\documentclass[fleqn,usenatbib]{mnras}

\usepackage[T1]{fontenc}
\usepackage{hyperref}
\usepackage{xcolor}

\DeclareRobustCommand{\VAN}[3]{#2}
\let\VANthebibliography\thebibliography
\def\thebibliography{\DeclareRobustCommand{\VAN}[3]{##3}\VANthebibliography}

\newcommand{\high}{\texttt{High Lensing}\xspace}
\newcommand{\low}{\texttt{Low Lensing}\xspace}

\usepackage{graphicx}	
\usepackage{amsmath}	
\usepackage{amssymb}	
\usepackage{xspace}
\usepackage{subfigure}

\newcommand{\Msun}{\,{\rm M_\odot}}

\newcommand{\angstrom}{\, \textup{\AA}}
\newcommand{\baymax}{\textsc{Baymax}\xspace}

\usepackage{newtxtext,newtxmath}

\title[No Evidence of Lensing in $z \gtrsim 6$ Chandra Quasars]{Lensing in the Darkness: A Bayesian Analysis of 22 Chandra Sources at $z \gtrsim 6$ Shows No Evidence of Lensing}

\author[F. Pacucci et al.]{
Fabio Pacucci$^{1,2}$\thanks{fabio.pacucci@cfa.harvard.edu},
Adi Foord$^{3}$\thanks{foord@stanford.edu},
Lucia Gordon$^{4}$ and
Abraham Loeb$^{1,2}$
\\
$^{1}$Center for Astrophysics $\vert$ Harvard \& Smithsonian,
Cambridge, MA 02138, USA\\
$^{2}$Black Hole Initiative, Harvard University,
Cambridge, MA 02138, USA\\
$^{3}$Kavli Institute of Particle Astrophysics and Cosmology, Stanford University, Stanford, CA 94305, USA\\
$^{4}$Department of Physics, Harvard University, Cambridge, MA 02138, USA\\
}

\date{\today}

\pubyear{2022}

\begin{document}
\label{firstpage}
\pagerange{\pageref{firstpage}--\pageref{lastpage}}
\maketitle

\begin{abstract}
More than $200$ quasars have been detected so far at $z > 6$, with only one showing clear signs of strong gravitational lensing. Some studies call for a missing population of lensed high-$z$ quasars, but their existence is still in doubt. A large fraction of high-$z$ quasars being lensed would have a significant effect on the shape of the intrinsic quasar luminosity function (QLF).
Here, we perform the first systematic search for lensed X-ray-detected quasars at $z \gtrsim 6$ employing a Bayesian analysis, with the code \baymax, to look for morphological evidence of multiple images that may escape a visual inspection.
We analyzed a sample of 22 quasars at $z > 5.8$ imaged by the Chandra X-ray observatory and found none with statistically significant multiple images. In the sub-sample of the 8 sources with photon counts $>20$ we exclude multiple images with separations $r>1\arcsec$ and count ratios $f>0.4$, or with separations as small as 0\farcs{7} and $f>0.7$ at $95\%$ confidence level. Comparing this non-detection with predictions from theoretical models suggesting a high and a low lensed fraction, we placed upper limits on the bright-end slope, $\beta$, of the QLF. Using only the sub-sample with 8 sources, we obtain, in the high-lensing model, a limit $\beta < 3.38$.
Assuming no multiple source is present in the full sample of 22 sources, we obtain $\beta < 2.89$ and $\beta < 3.53$ in the high and low lensing models, respectively.
These constraints strongly disfavor steep QLF shapes previously proposed in the literature.
\end{abstract}




\section{Introduction} \label{sec:intro}
Gravitational lensing alters the appearance of background astrophysical objects due to the presence of a mass distribution along the line of sight (see, e.g., the review by \citealt{Schneider_1992}). In strong gravitational lensing (see, e.g., \citealt{Treu_2010}), a foreground object with a sufficiently large surface mass density distorts the light from a background source such as a quasar, leading to multiple images and a magnification of its brightness. Gravitational lensing is a powerful tool at the astronomer's disposal, used to study the distribution of dark matter in clusters (e.g., \citealt{Natarajan_2017_lensing}) and sub-structures (e.g., \citealt{Meneghetti_2020}), to detect high-redshift sources otherwise too faint (e.g., \citealt{Fan_2019}), to investigate the contribution of compact objects to dark matter (e.g., \citealt{Alcock_1996}), and to constrain cosmological parameters (e.g., \citealt{holicow_2020}).

In this study, we focus on the effect of gravitational lensing on quasars \citep{Turner_1980}, i.e. supermassive black holes that are accreting at large rates and luminosities \citep{Schmidt_1968}. Nowadays, tens of thousands of quasars have been discovered via systematic surveys (e.g., the Sloan Digital Sky Survey, \citealt{Shen_2011}), even at very high-$z$ (e.g., \citealt{Fan_2006, Mortlock_2011, Wang_2021}). The detailed study of mass and redshift distribution of quasars is relevant for a number of cosmological implications, including the origin of the first population of black holes (e.g., \citealt{Haiman_Quataert_2004, Volonteri_2010, Haiman_2013, Gallerani_2017,   Woods_2019, Inayoshi_review_2019, Pacucci_2022}).

Although $> 200$ quasars are now detected at $z\sim 6$, remarkably only one lensed quasar at $z > 6$ was discovered \citep{Fan_2019}: J0439+1634 at $z=6.51$, with a possible magnification factor $\sim 50$.
Quasars at $z > 6$ are peculiar, as their emission at wavelengths shorter than that of the Lyman-$\alpha$ line ($1216 \angstrom$) is absorbed by intervening neutral matter \citep{Gunn_Peterson_1965}.
The fact that only one lensed quasar at $z > 6$ was discovered so far is interesting. \cite{Wyithe_Loeb_2002} predicted that up to $1/3$ of quasars at $z > 6$ should be strongly magnified (by a magnification factor $\mu > 10$), due to the presence of galaxies along the line of sight.
The lensing optical depth \citep{Turner_1980, Hilbert_2007}, $\tau_L$, is a measure of the probability that a source at redshift $z_s$ is lensed by a lens at $z_l < z_s$. Intuitively, $\tau_L$ increases with $z_s$, as there are more potential lensing galaxies along the line of sight. 

The discovery of the $z > 6$ lensed quasar J0439+1634 prompted a deeper look into selection criteria. \cite{Fan_2019} argued that standard techniques based on the non-detection of the quasar at wavelengths shorter than the Lyman-$\alpha$ may cause a selection bias against lensed quasars. Additionally, multiply-imaged quasars at $z>6$ might not be recognized as such due to the limited angular resolution of current telescopes \citep{Pacucci_Loeb_2019}. Recently, \cite{Yue_2021} revised the lensed fraction calculated by previous works, suggesting that improved accounting of the galaxy velocity dispersion functions would reduce the lensing probability by a factor of $\sim 10$, shrinking the gap between theoretical predictions and actual observations of the lensed fraction.

Understanding the extent of the lensed population of $z>6$ quasars is fundamental, as a large lensing fraction has profound effects on our measurements of the quasar luminosity function (QLF) and of the quasar mass function (e.g., \citealt{Wyithe_Loeb_2002, Comerford_2002}). A large lensed fraction would cause a significant departure of the observed QLF from the intrinsic one, especially at its bright-end. Regarding the mass function, \cite{Fujimoto_2020} claimed that the most massive quasar found at $z > 6$, with a mass $1.2 \times 10^{10} \Msun$ \citep{Wu_2015}, could be extremely magnified, and \cite{Pacucci_Loeb_2019_mirage} suggested that this would hint at a very large fraction of lensed high-$z$ quasars. Further studies of the quasar proximity zone of this source ruled out extreme magnification \citep{Davies_2020}. Additionally, the recent discovery of a putative physical pair of quasars at $z = 5.66$, separated by only $7.3$ kpc \citep{Yue_2021_pair} makes it even more urgent to better understand the role of lensing at $z\gtrsim 6$. 

The value of the bright-end slope of the QLF is a powerful indicator of the lensing probability: larger values, corresponding to a steeper bright-end, would lead to a greater probability of quasars being lensed \citep{Pei_1993, Pei_1995, Comerford_2002, Keeton_2005}. \cite{Pacucci_Loeb_2019} suggested that the probability of observing a gravitationally lensed quasar with magnification $\mu < 10$ is higher than the observed frequency of detections among the entire sample of $z>6$ quasars.

In this study, we use Chandra X-ray data for $22$ $z\gtrsim 5.8$ quasars to investigate the possibility that they are lensed, employing a Bayesian analysis performed with \baymax \citep{Foord_2019}.
In \S \ref{sec:theory} we describe the gravitational lensing models employed, as well as \baymax. In \S \ref{sec:data} we describe the sample of Chandra sources used and the limitations of their analysis. Finally, in \S \ref{sec:results} and \S \ref{sec:disc_concl} we present our results and conclusions.

\section{METHODOLOGY} 
\label{sec:theory}
This Section briefly describes the lensing models adopted as baseline for this work. The interested reader is referred to \cite{Pacucci_Loeb_2019} and \cite{Yue_2021} for a more in-depth description.
Additionally, we also describe \baymax, and the reader is referred to \cite{Foord_2019, Foord_2020} for a complete review of the code.

\subsection{Gravitational Lensing Models} 
\label{subsec:lensing}
In this study we consider two different lensing models. The first, introduced by \cite{Pacucci_Loeb_2019}, predicts a large fraction of lensed quasars at $z\sim 6$ and is in accordance with previous theoretical estimates (e.g., \citealt{Wyithe_Loeb_2002, Comerford_2002}). For example, this model predicts a fraction of $\sim 20\%$ lensed quasars at $z\sim 6$, assuming a value $\beta = 3.6$ of the bright-end slope of the QLF. 
The second model, recently introduced by \cite{Yue_2021} and in accordance with current observations, predicts a much lower fraction of lensed quasars at high-$z$. For comparison, this model predicts a fraction of $\sim 4\%$ lensed quasars at $z\sim 6$, again assuming $\beta = 3.6$. These two models differ significantly in the assumptions on the velocity dispersion of foreground galaxies, which act as lenses for the high-$z$ quasars.

In order to easily distinguish between the two models, throughout the remaining text we use the label \high to describe the model in \cite{Pacucci_Loeb_2019}, and \low to describe the model in \cite{Yue_2021}.

The interested reader is referred to the very broad descriptions of the models presented in \cite{Pacucci_Loeb_2019} and \cite{Yue_2021} to better grasp the significance of the parameter $\beta$ and its influence on the magnification bias and, hence, on the lensed fraction.

\subsection{BAYMAX} 
\label{subsec:baymax}
This study makes use of \baymax \citep{Foord_2019}, a code originally developed to analyze Chandra X-ray observations of Active Galactic Nuclei (AGN) and compute the likelihood that an object is better described by a single or a dual source model. While in the original interpretation of the results provided by \baymax the AGN would be a physical double system, with two super-massive black holes orbiting in the core of their merged galaxies, in our current interpretation the multiple images would be caused by gravitational lensing. Remarkably, the code is able to assign model probabilities for the source independently of the nature of the mechanism causing the multiple image. \baymax employs a Bayesian approach to calculate the best model to describe the source, once the Chandra's Point Spread Function (PSF) is properly taken into account.

Following the classic Bayesian statistical approach (see, e.g., \citealt{Jeffreys_1935}), we name $D$ the available data. In our case, $D$ is the distribution of photons of the AGN in the Chandra image, once it is corrected for PSF effects. Hence, the conditional probability of observing the data $D$ given the model with a single source is indicated as $P(D|M_1)$ and, similarly, the conditional probability of observing the data $D$ given the model with a double source is $P(D|M_2)$. Of course, both the single-source and the double-source models depend on additional parameters, which we generally indicate as $\theta_1$ for the single source and $\theta_2$ for the double source. In our case, these parameters are the angular distance between the two sources ($r$) and the count ratio ($f$, which represents the ratio between the number of counts associated with the secondary point source and the number of counts associated with the primary point source).
Generally speaking, a larger angular separation between the two sources, as well as a count ratio near unity (i.e., equally bright sources) is more likely to result in a detection. In fact, if a source is gravitationally lensed, secondary X-ray images below a certain flux cannot be detected.

The Bayes factor ($\mathcal{BF}$), which is our primary indicator of the preference for one model or the other, is calculated by marginalizing over the entire parameter space of the models. Following \cite{Foord_2019}, the Bayes factor is then calculated as:
\begin{equation}
    \mathcal{BF} = \frac{P(D|M_2, \theta_2) \, P(\theta_2, M_2) \mathrm{d}\theta_2}{P(D|M_1, \theta_1) \, P(\theta_1, M_1) \mathrm{d}\theta_1} \, .
\end{equation}

As the Bayes factor is defined as the marginal probability of the model with a double source over the marginal probability of the model with a single source, $\ln{\mathcal{BF}} > 0$ indicates a preference for the double source model. \baymax calculates the Bayes factor using nested sampling \citep{Skilling_2004} via the {\tt python} packakge {\tt nestle} \footnote{https://github.com/kbarbary/nestle} and uses {\tt PyMC3} \citep{Salvatier_2016} for parameter estimation. We refer the reader to \cite{Foord_2019, Foord_2020} for a detailed discussion of the statistical techniques used to estimate likelihoods and posterior densities.
Also, we note that \baymax is specifically designed to search for double images in Chandra data, while gravitational lensing can also produce quadrupole, or even higher order, images, although they are rarer than doubles. \baymax is generally able to detect multiplicity also in the case of a number $>2$ of images.

The information provided by \baymax can be a strong indication of the multiplicity of a Chandra-detected source. In case of a suspected multiple source, this needs to be complemented by additional data to confirm gravitational lensing. First, higher resolution imaging, e.g. with the Hubble Space Telescope (HST) or the Atacama Large Millimeter/submillimeter Array (ALMA), could provide important information regarding the location of the multiple images. Also, a detection of a foreground object in the same line of sight would support the lensing hypothesis (see, e.g., \citealt{Fan_2019}). The study of the properties of the lens would help in the definition of a lensing model that accounts for the relative magnifications and locations of the multiple images. Finally, the study of the quasar proximity zone and the derivation of the black hole mass from spectral lines could provide independent data on black hole growth rate, thus informing the lensing hypothesis (see, e.g., \citealt{Davies_2020}).

\section{DATA}
\label{sec:data}
In this Section we describe the Chandra data used in this study.

\begin{table}
\begin{center}
        \begin{tabular}{ || c | c| c |}
 \hline
 \textbf{Source name} & \textbf{Chandra Obs. ID} & \textbf{Exposure Time (s)} \\
 \hline
 
J002429+391318 & 20416 &  19700 \\

J084035+562419 & 05613 &  15840 \\
 
J022601+030259 & 20390 &  25900 \\

J133550+353315 & 07783 &  23470 \\

J111033-132945 & 20397 &  59330 \\
\dots & 22523 & 42840 \\
\dots & 22523 & 42840 \\
\dots & 23199 & 40630 \\
\dots & 23153 & 23990 \\
\dots & 23018 & 9980 \\
 
J150941-174926 & 20391 &  26760 \\
 
J162331+311200 & 05607 &  17210 \\
 
J223255+293032 & 20395 & 54210 \\

J005006+344522 & 20393 &  33490 \\

J152637-205000 & 22233 &  39670 \\
\dots & 22165 &  32570 \\
\dots & 22231 &  29670 \\
\dots & 22232 &  21780 \\
\dots & 20469 &  16830 \\

J141111+121737 & 05611 & 14270 \\

J010013+290225 & 17087 &  14800\\
 
J163033+401209 & 05618 & 27390 \\

J213233+121755 & 20417 &  17820 \\

J000552-000655 & 05617 & 16930 \\

J083643+005453 & 03359 &  5680\\

J160253+422824 & 05609 &  13200 \\

J114816+525150 & 17127 & 77770\\
 
J164121+375520 & 21961 &  33480 \\
\dots & 20396 &  20830 \\

J203209-211402 & 21725 & 73370 \\
\dots & 20470 & 44480 \\
\dots & 21726 & 33070 \\

J103027+052455 & 19987 & 126380 \\
\dots & 20045 & 61270 \\
\dots & 19926 & 49420 \\
\dots & 18185 & 46330\\
\dots & 18187 & 40400 \\
\dots & 20046 & 36590 \\
\dots & 18186 & 34620\\
\dots & 19994 & 32650 \\
\dots & 19995 & 26720\\
\dots & 20081 & 24910 \\
\dots & 03357 & 7950\\

J130608+035626 & 03358 &  118240\\
\dots & 03966 &  8160  \\
\hline
        \end{tabular}
    \end{center}
\caption{Summary of the Chandra data of the 22 $z \gtrsim 5.8$ sources studied. From left to right: the source name, Chandra observation ID, and exposure time of Chandra observation.}
\label{table:cxoinfo}
\end{table}

\begin{table*}
\begin{center}
        \begin{tabular}{ || c | c| c | c | c | c | c | }
 \hline
 \textbf{Source name} & \textbf{$\alpha$} & \textbf{$\delta$} & \textbf{Redshift}  & \textbf{X-ray Counts} & \textbf{$\ln{\mathcal{BF}}$} & \textbf{Reference} \\
 \hline
 
J002429+391318 & 00:24:29.77 & +39:13:18.98 & 6.621 & 3 & $0.12\pm0.25$ & Tang et al. (2017) \\

J084035+562419 & 08:40:35.09 & +56:24:19.90 & 5.816 & 4 & $0.15\pm0.31$ & Fan et al. (2006) \\
 
J022601+030259 & 02:26:01.88 & +03:02:59.40 & 6.541 & 5 & $0.61\pm0.23$ & Venemans et al. (2015) \\

J133550+353315 & 13:35:50.806 & +35:33:15.80 & 5.901 & 6 & $0.13\pm0.27$ & Fan et al. (2006) \\

J111033-132945 & 11:10:33.98 & -13:29:45.60 & 6.515 & 6 & $0.18\pm0.49$ & Venemans et al. (2015) \\

J150941-174926 & 15:09:41.78 & -17:49:26.80 & 6.123 & 6 & $-0.01\pm0.29$ & Willott et al. (2007) \\
 
J162331+311200 & 16:23:31.81 & +31:12:00.50 & 6.254 & 7 & $0.11\pm0.27$ & Fan et al. (2004) \\
 
J223255+293032 & 22:32:55.15 & +29:30:32.20 & 6.666 & 8 & $0.07\pm0.27$ & Venemans et al. (2015)\\

J005006+344522 & 00:50:06.67 & +34:45:22.60 & 6.251 & 9 & $0.21\pm0.32$ & Willott et al. (2010) \\

J152637-205000 & 15:26:37.84 & -20:50:00.80 & 6.586 & 14 & $0.43\pm0.47$ & Mazzucchelli et al. (2017) \\
 
J141111+121737 & 14:11:11.29 & +12:17:37.4 & 5.854 & 13 & $-0.07\pm0.36$ & Fan et al. (2004) \\

J010013+290225 & 01:00:13.02 & +28:02:25.80 & 6.327 & 14 & $-0.15\pm0.37$  & Wu et al. (2015) \\
 
J163033+401209 & 16:30:33.90 & +40:12:09.60 & 6.066 & 14 & $-0.01\pm0.32$ & Fan et al. (2003) \\

J213233+121755 & 21:32:33.19 & +12:17:55.46 & 6.588 & 15 & $-0.06\pm0.31$ & Mazzucchelli et al. (2017) \\
 
 \hline

J000552-000655 & 00:05:52.34 & -00:06:55.80 & 5.844 & 21 & $0.07\pm0.26$ & Fan et al. (2004) \\

J083643+005453 & 08:36:43.86 & +00:54:53.23 & 5.834 & 24 & $-0.18\pm0.33$ & Fan et al. (2001) \\

J160253+422824 & 16:02:54.18 & +42:28:22.90 & 6.083 & 27 & $-0.22\pm0.39$ & Fan et al. (2004) \\

J114816+525150 & 11:48:16.65 & +52:51:50.21 & 6.419 & 35 & $-0.04\pm0.34$ & Fan et al. (2003) \\
 
J164121+375520 & 16:41:21.64 & +37:55:20.50 & 6.047 & 45 & $0.53\pm0.41$ & Willott et al. (2007) \\
 
J203209-211402 & 20:32:09.99 & -21:14:02.31 & 6.234 & 72 & $-0.77\pm0.48$ & Banados et al. (2016) \\

\hline

J103027+052455 & 10:30:27.10 & +05:24:55.00 & 6.280 & 132.325 & $0.31\pm0.66$ & Fan et al. (2001) \\

J130608+035626 & 13:06:08.26 & +03:56:26.35 & 6.034 & 132 & $0.42\pm0.43$ & Fan et al. (2001) \\
\hline
        \end{tabular}
    \end{center}
\caption{Summary of the properties of the 22 $z \gtrsim 5.8$ Chandra sources studied. From left to right: the source name, redshift, background-subtracted quasar counts ($0.5-8$ keV) across all observations, computed Bayes factor, and original reference for the source. We denote the 3 count bins used for our sensitivity tests with horizontal line breaks (see \ref{subsec:sensitivity}).}
\label{table:sources}
\end{table*}

\subsection{Chandra quasar sample}
\label{subsec:chandra_data}
We use a sample of $z > 5.8$ quasars reported by \cite{Li_2021}. In particular, they analyze $152$ quasars, including $46$ with a $>3\sigma$ detection by Chandra. Of these $152$ quasars, 76 have $z>5$, 35 have $z>6$, and 2 have $z>7$. 

We restrict our analysis to Chandra observations with $z>5.8$, with off-axis angles below 1$\arcmin$ and non-zero values for the background-subtracted quasar counts in the $0.5-8$ keV energy range. The Chandra PSF becomes increasingly difficult to model as a function of increasing off-axis angles, and thus \baymax is most powerful (and sensitive to smaller separations and count ratios) when analysing on-axis observations. This results in 22 sources, where 16 have a single Chandra observation, 2 have 2 observations, 1 has 3 observations, 2 have 5 observations, and 1 has 11 observations. In Table \ref{table:cxoinfo} we list the observations used in our analysis for each source.

The \cite{Li_2021} sample was chosen for the uniformity in its data analysis of Chandra sources, and because it provided a consistent catalog of high-$z$ quasars. Table \ref{table:sources} includes the main descriptors of the data used, including the spectroscopic redshift, the net number of quasar counts in the energy range $0.5-8$ kev, as well as the reference to the discovery paper of each source \citep{Fan_2001, Fan_2003, Fan_2004, Fan_2006, Willott_2007, Willott_2010, Venemans_2015, Wu_2015, Banados_2016, Tang_2017, Mazzucchelli_2017}.

The angular resolution of the Chandra X-ray observatory is nominally $\sim 0\farcs{5}$, which is significantly worse than the HST  with $\sim 0\farcs{1}$. In fact, the final confirmation that the $z = 6.51$ quasar J0439+1634 is lensed came from a high-resolution image obtained with the HST \citep{Fan_2019}. The rationale for our choice of using Chandra sources for a search of lensed quasars is two pronged:
\begin{enumerate}
    \item X-ray detected sources at high-$z$ are most likely associated with quasar activity, i.e. with accretion on a super-massive black hole.
    \item \baymax adopts photon statistics and PSF analysis specifically designed for Chandra. The complex Bayesian analysis of Chandra images performed by \baymax allows the identification of multiple images of a source that, by eye inspection, looks single.
\end{enumerate}

\cite{Yue_2021} calculate the distribution of separations between multiple images in their model, obtaining that $\sim 85\%$ of lensed quasars at $z = 6$ should have angular separations $> 0\farcs{5}$, and hence larger than the angular resolution of Chandra. Note that this is comparable to the fraction calculated in \cite{Pacucci_Loeb_2019} at $z=6.5$. However, the detectability of any putative lensed system via \baymax will strongly rely on the total number of counts, the separation, and the count ratio (e.g., below $\sim$10 counts, it is difficult to find dual X-ray point sources with \baymax at high significance for most of parameter space). In Section~\ref{subsec:sensitivity} we review our sensitivity in parameter space for the sample, as a function of their total counts.

\begin{figure*}
     \centering
     \begin{subfigure}
         \centering
         \includegraphics[width=0.3\textwidth]{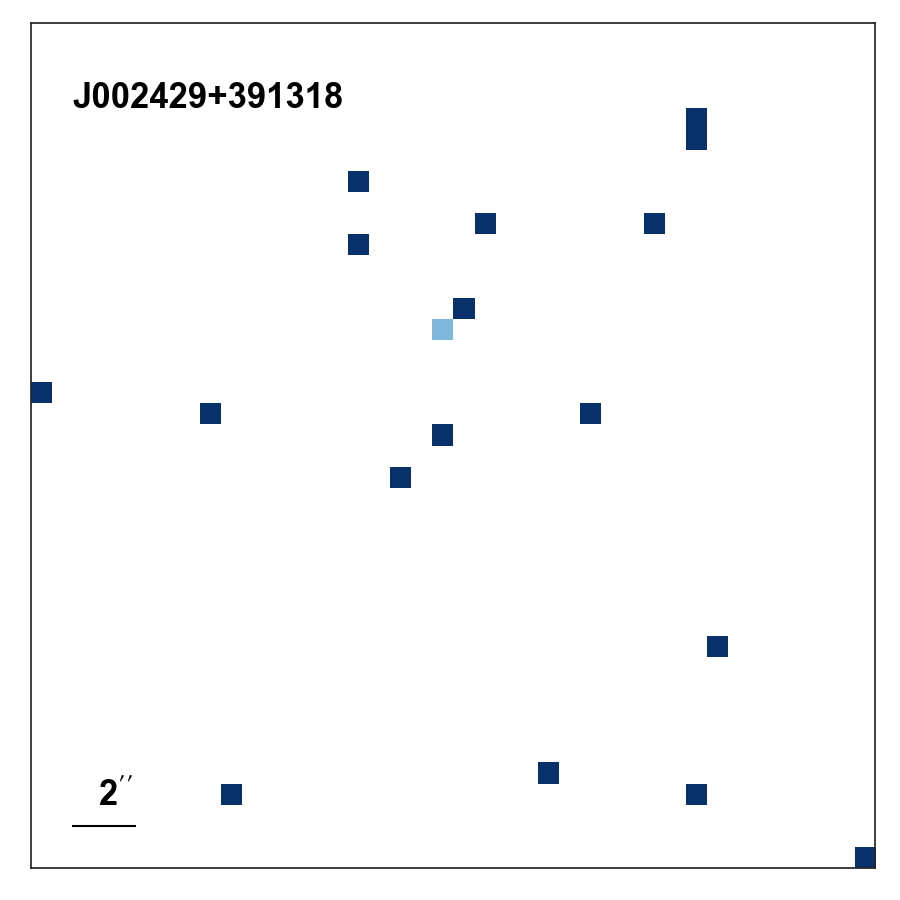}
     \end{subfigure}
     \hfill
     \begin{subfigure}
         \centering
         \includegraphics[width=0.3\textwidth]{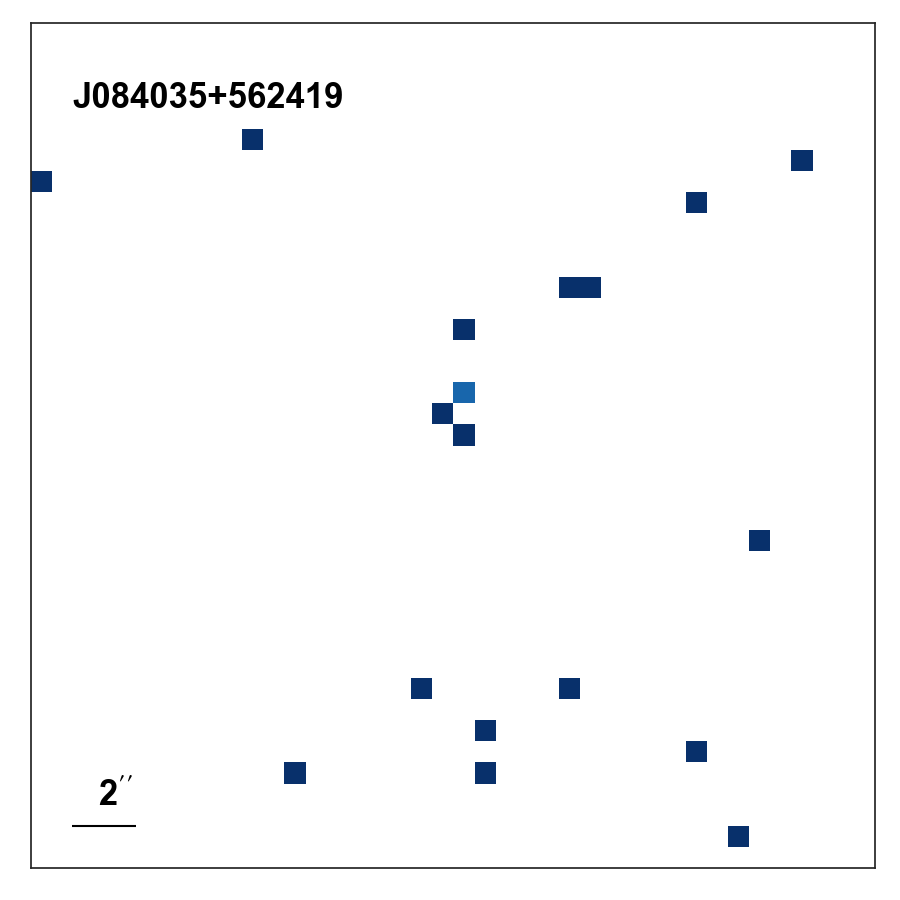}
     \end{subfigure}
     \hfill
     \begin{subfigure}
         \centering
         \includegraphics[width=0.3\textwidth]{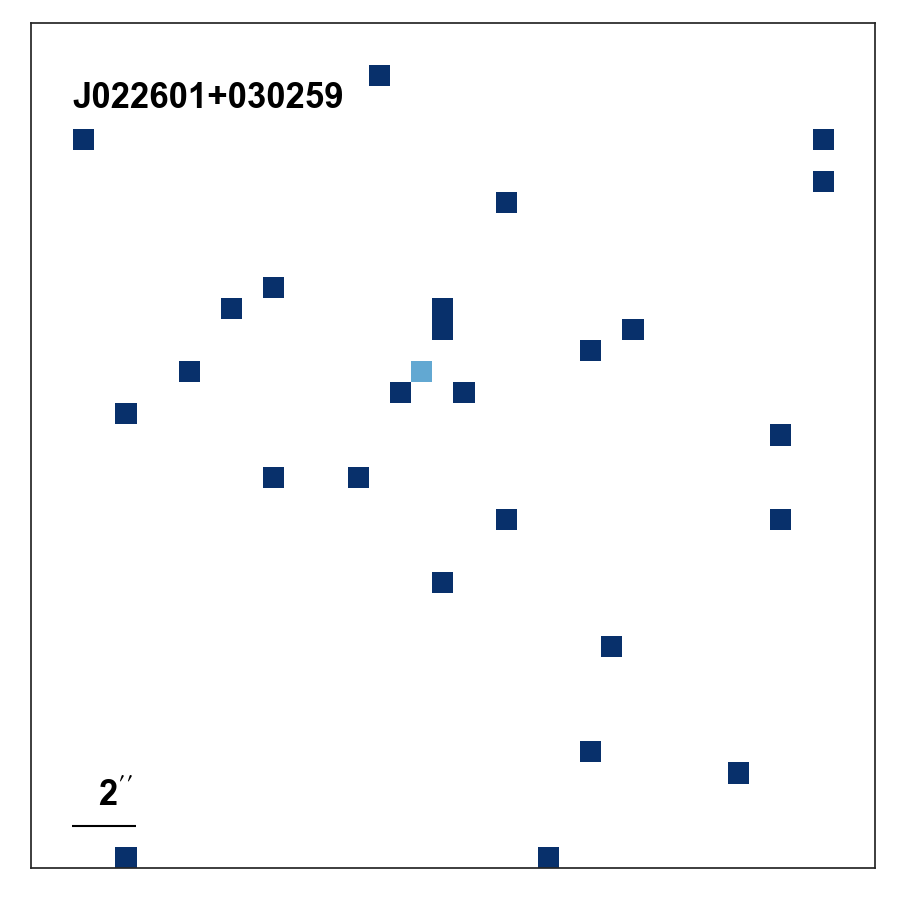}
     \end{subfigure}
     
     \vspace{0.2cm}
     
    \begin{subfigure}
         \centering
         \includegraphics[width=0.3\textwidth]{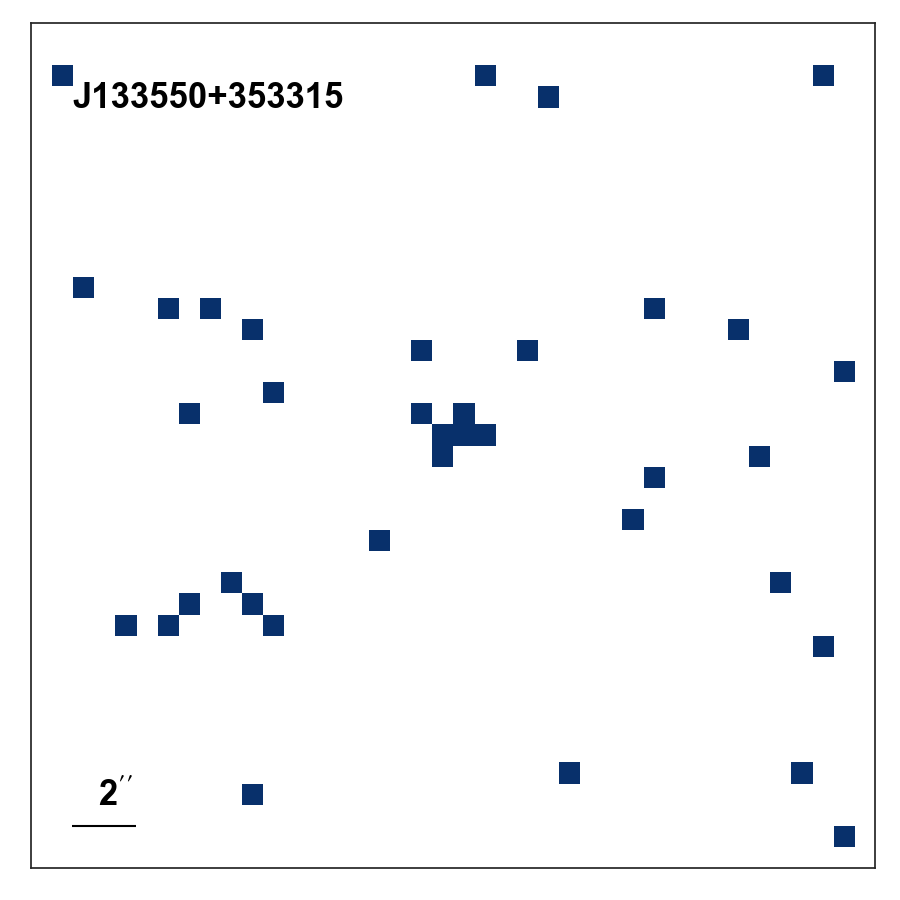}
     \end{subfigure}
     \hfill
     \begin{subfigure}
         \centering
         \includegraphics[width=0.3\textwidth]{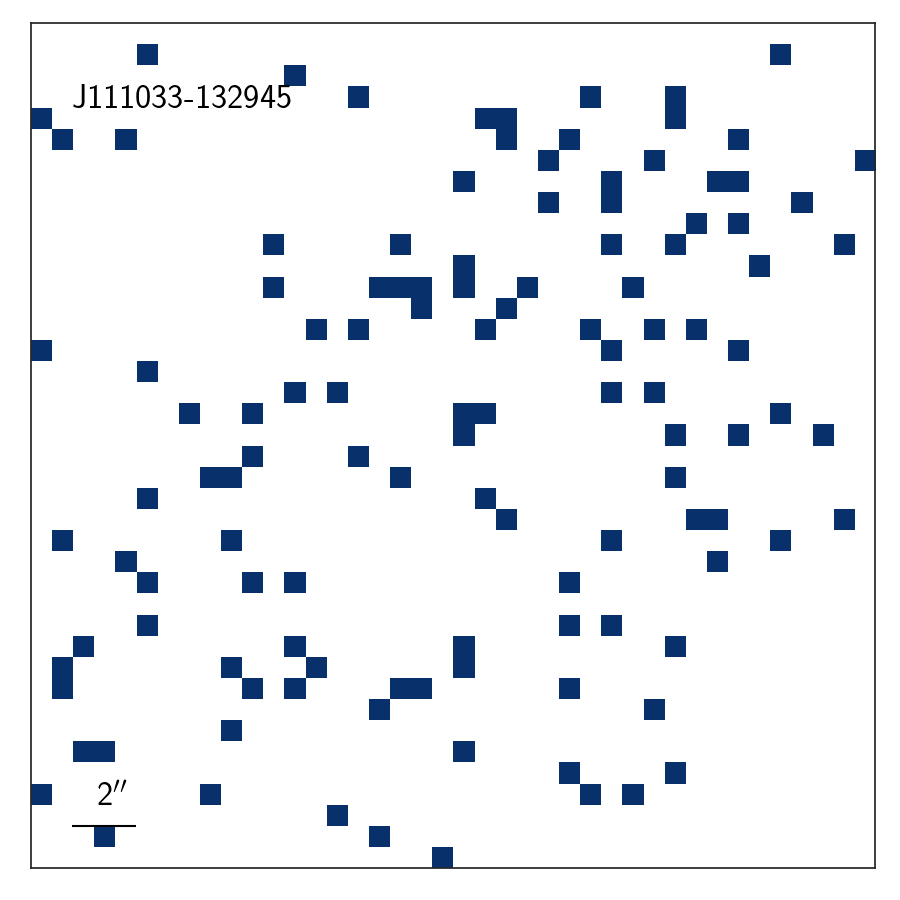}
     \end{subfigure}
     \hfill
     \begin{subfigure}
         \centering
         \includegraphics[width=0.3\textwidth]{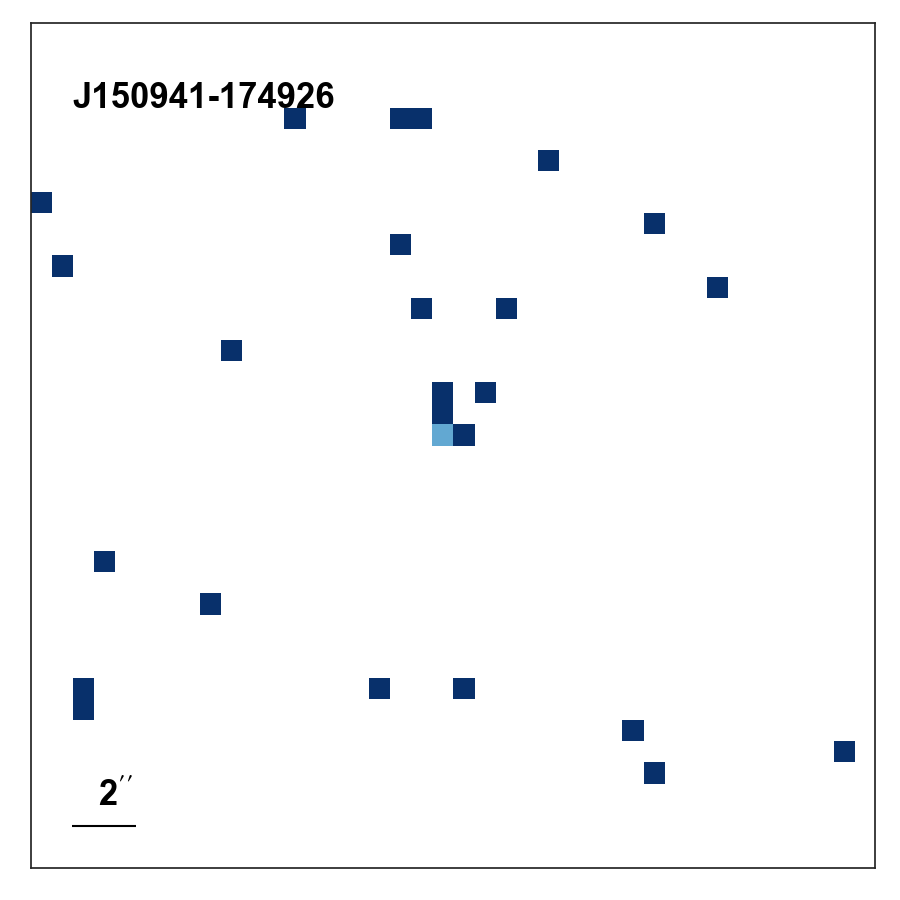}
     \end{subfigure}
     
     \vspace{0.2cm}
     
     \begin{subfigure}
         \centering
         \includegraphics[width=0.3\textwidth]{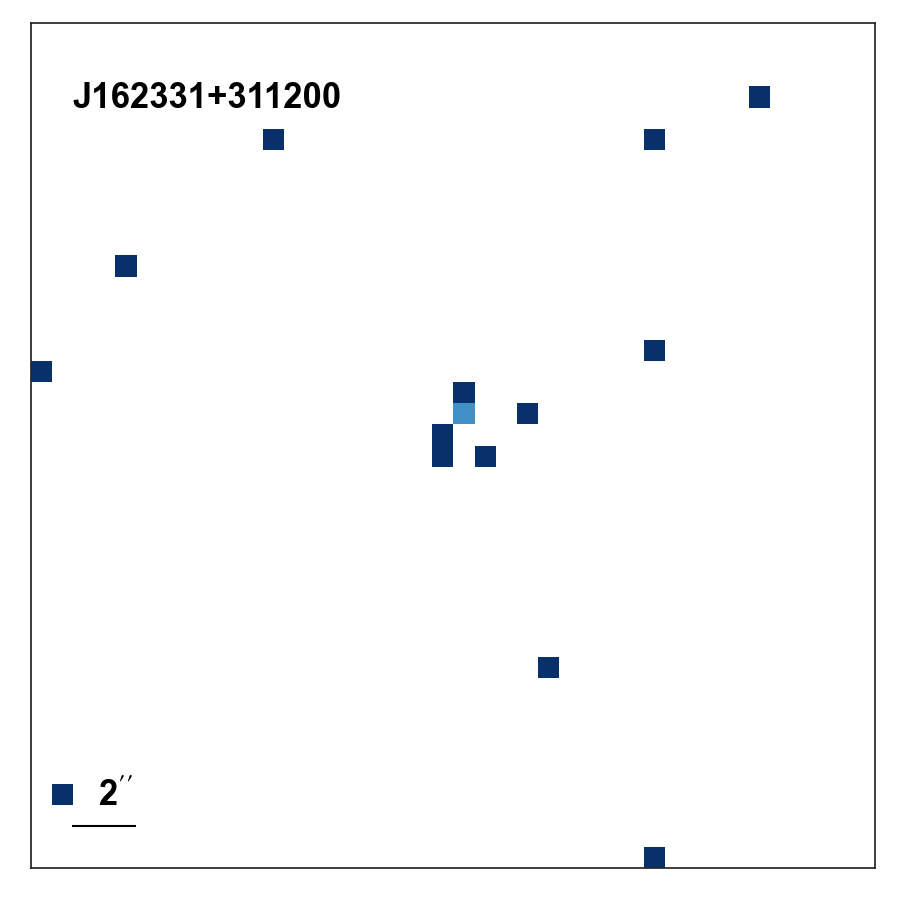}
     \end{subfigure}
     \hfill
     \begin{subfigure}
         \centering
         \includegraphics[width=0.3\textwidth]{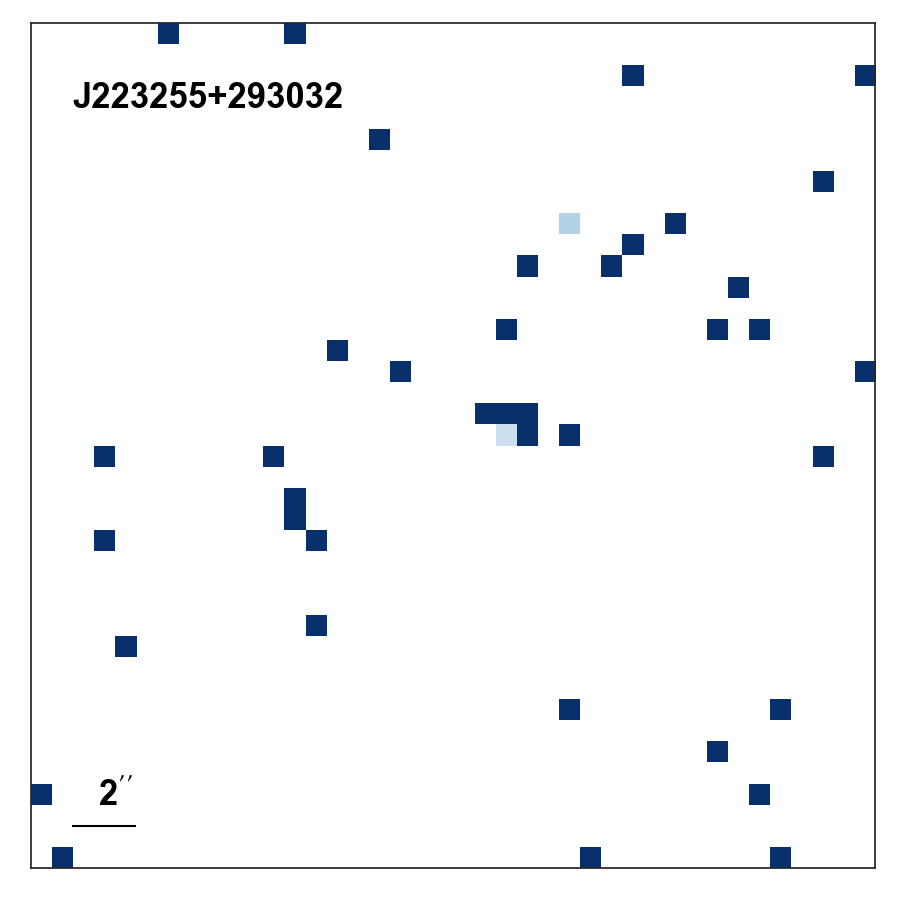}
     \end{subfigure}
     \hfill
     \begin{subfigure}
         \centering
         \includegraphics[width=0.3\textwidth]{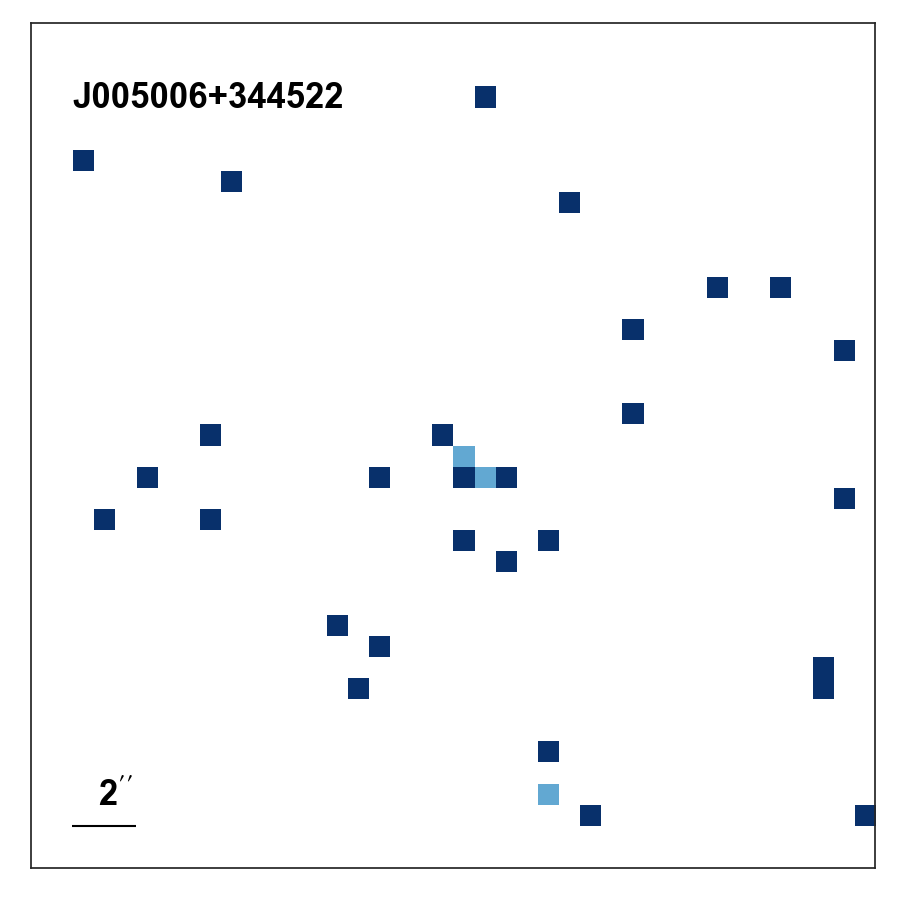}
     \end{subfigure}
          \hfill
          
    \vspace{0.2cm}
    
    \begin{subfigure}
         \centering
         \includegraphics[width=0.3\textwidth]{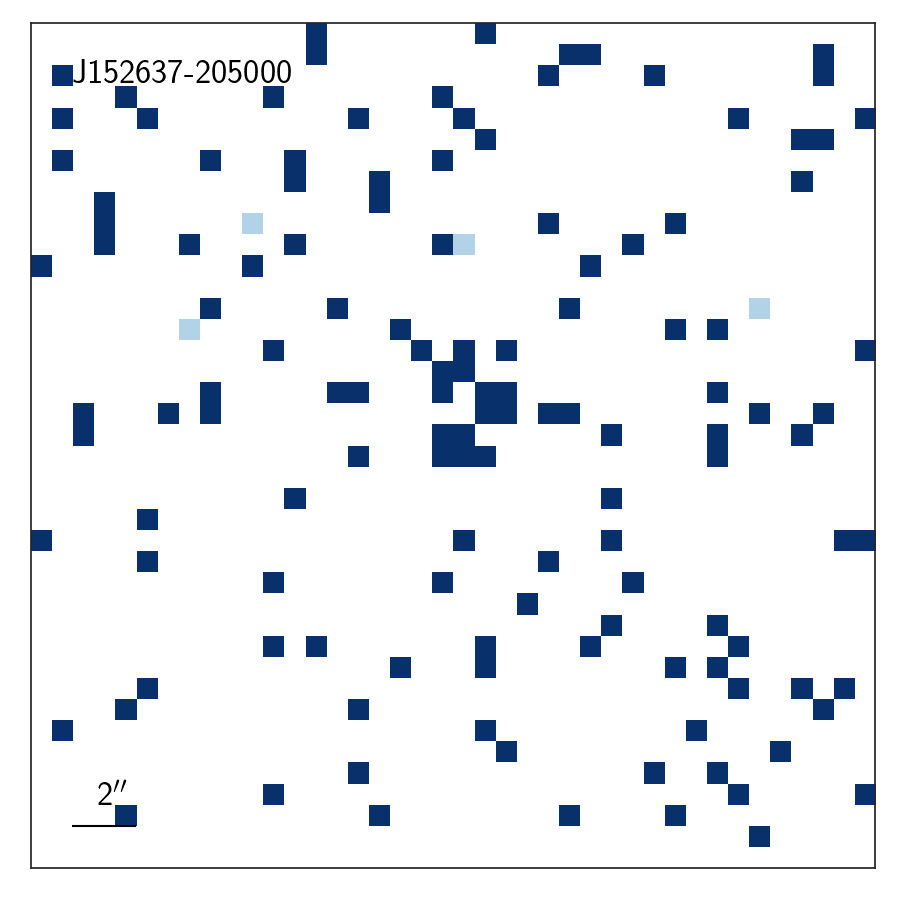}
    \end{subfigure}
    \hfill
    \begin{subfigure}
         \centering
         \includegraphics[width=0.3\textwidth]{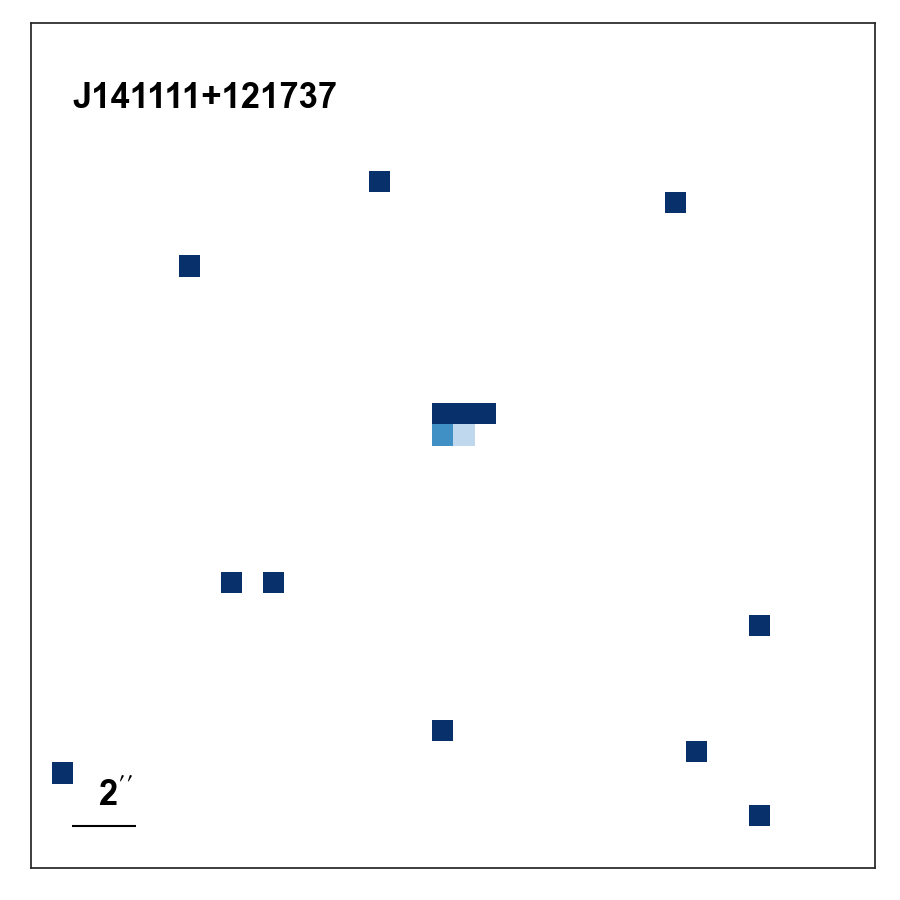}
    \end{subfigure}
    \hfill
    \begin{subfigure}
         \centering
         \includegraphics[width=0.3\textwidth]{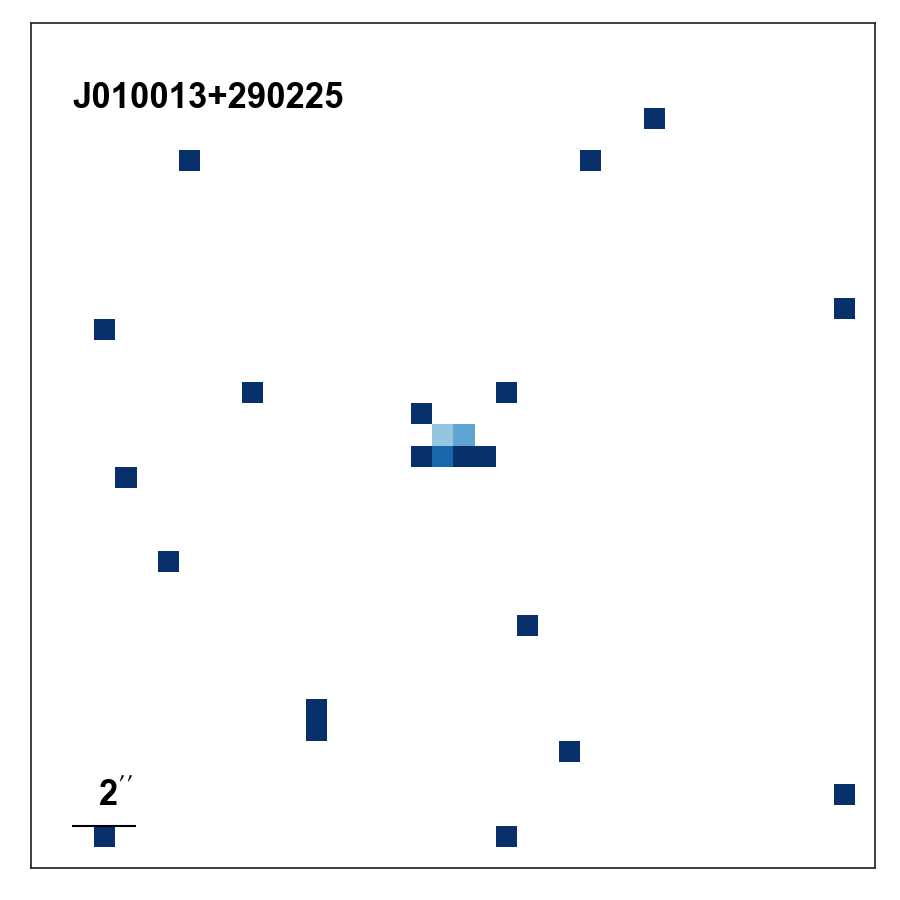}
     \end{subfigure}
     
        \caption{Binned X-ray images of selected sources in our dataset. Photons have energies in the range $0.5-8$ keV, and the scale is $0.5$\arcsec per pixel.}
        \label{fig:X_images}
\end{figure*}

\begin{figure*}
     \centering

    \begin{subfigure}
         \centering
         \includegraphics[width=0.3\textwidth]{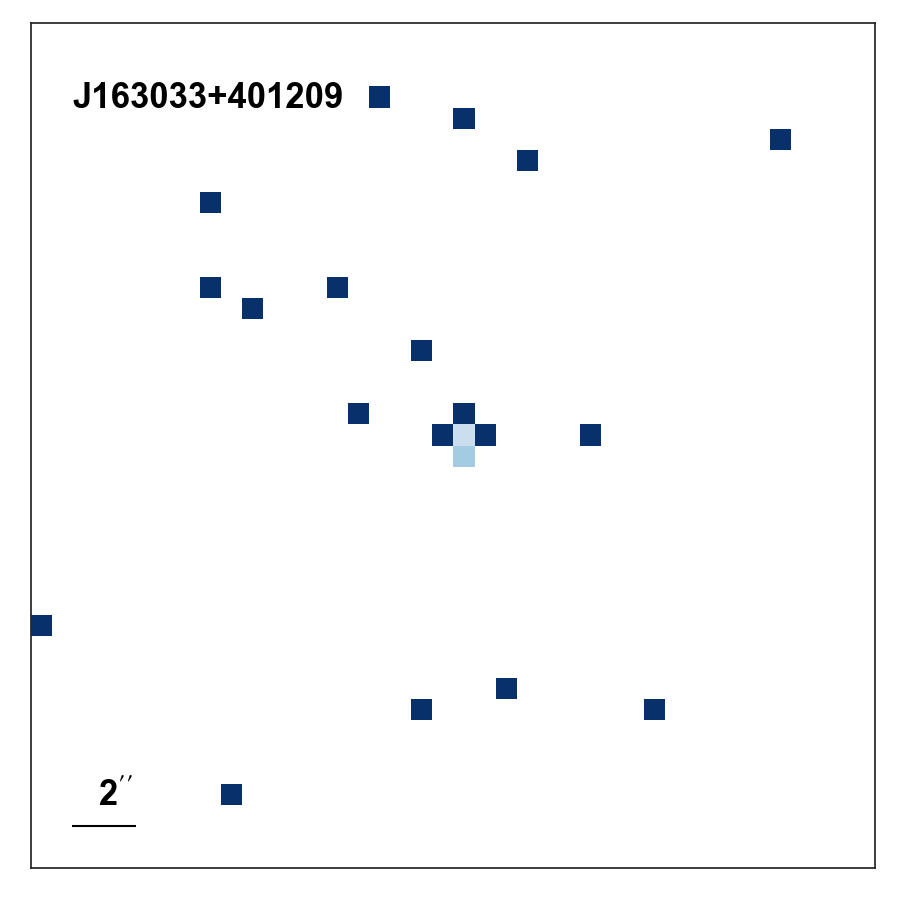}
    \end{subfigure}
    \hfill
    \begin{subfigure}
         \centering
         \includegraphics[width=0.3\textwidth]{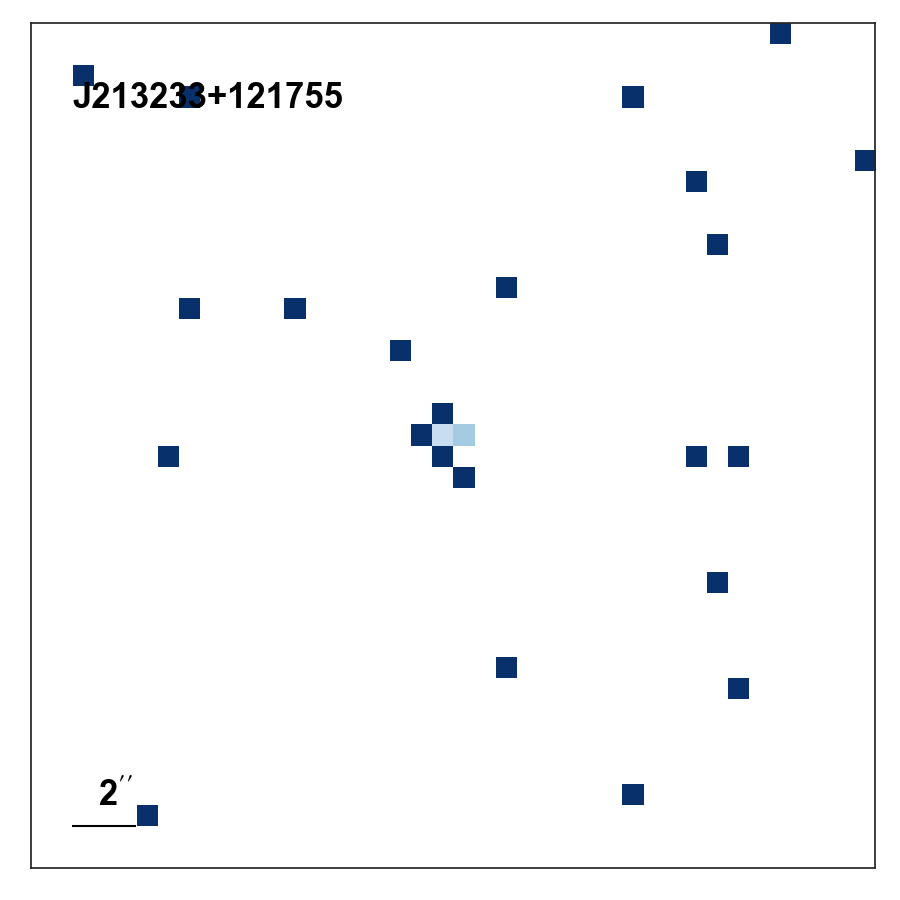}
    \end{subfigure}
    \hfill
    \begin{subfigure}
         \centering
         \includegraphics[width=0.3\textwidth]{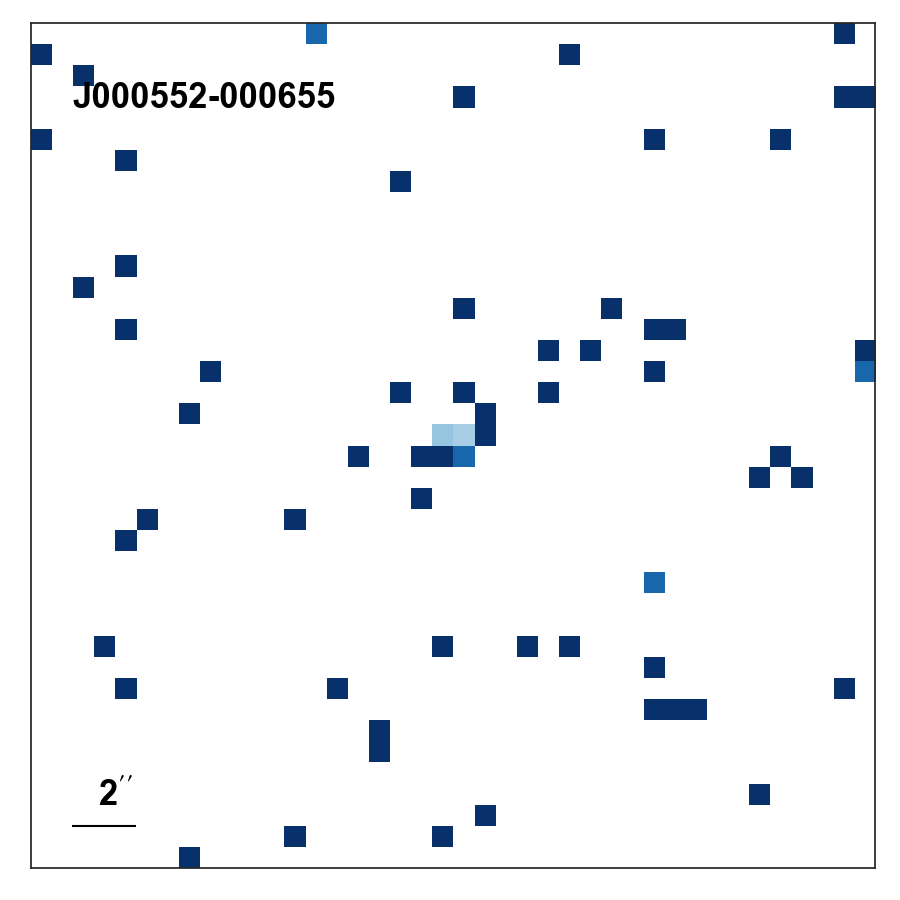}
     \end{subfigure}
     
    \vspace{0.2cm} 
     
    \begin{subfigure}
         \centering
         \includegraphics[width=0.3\textwidth]{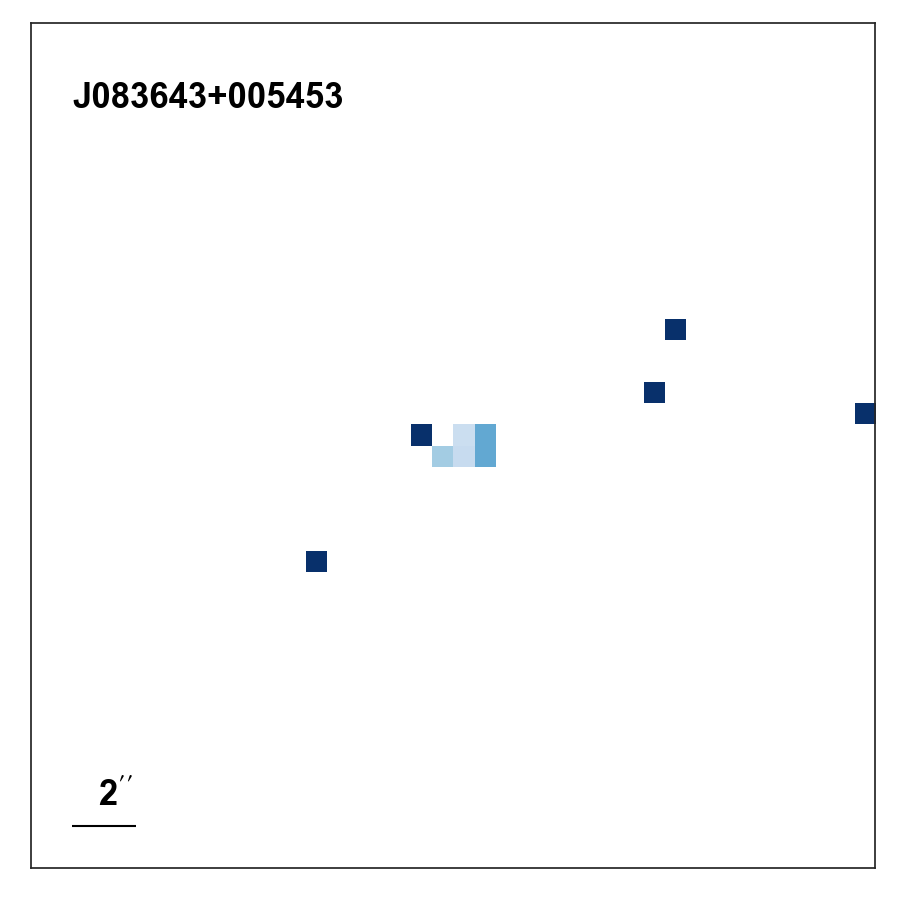}
    \end{subfigure}
    \hfill
    \begin{subfigure}
         \centering
         \includegraphics[width=0.3\textwidth]{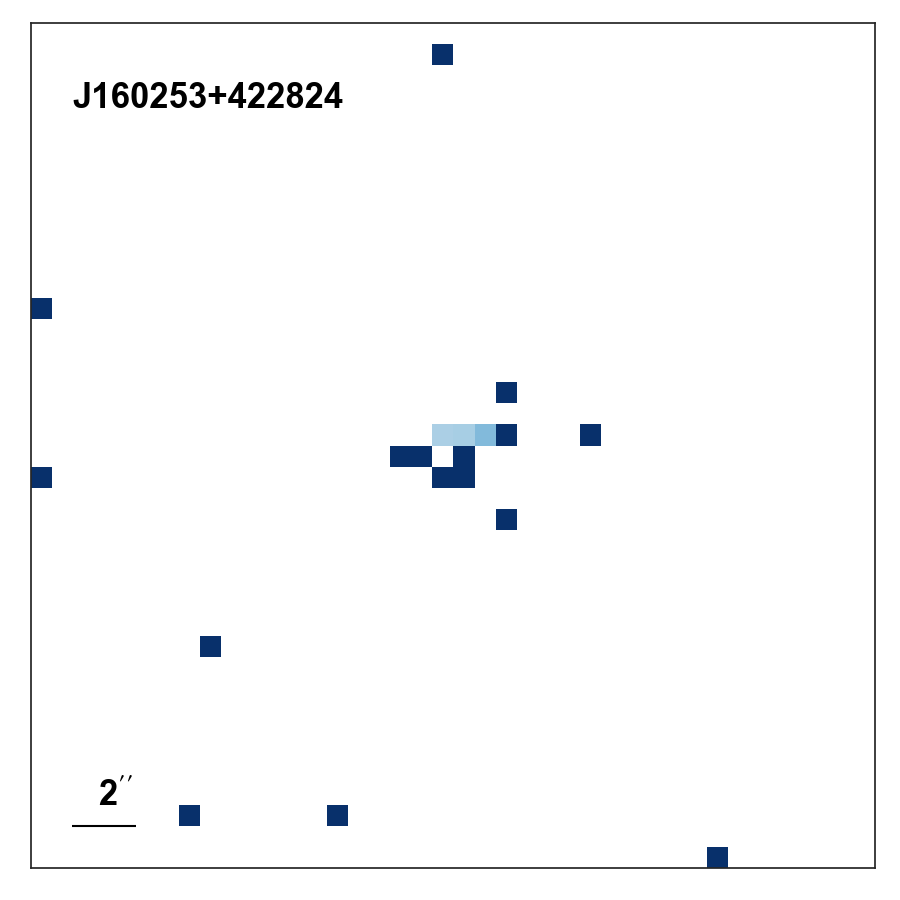}
    \end{subfigure}
    \hfill
    \begin{subfigure}
         \centering
         \includegraphics[width=0.3\textwidth]{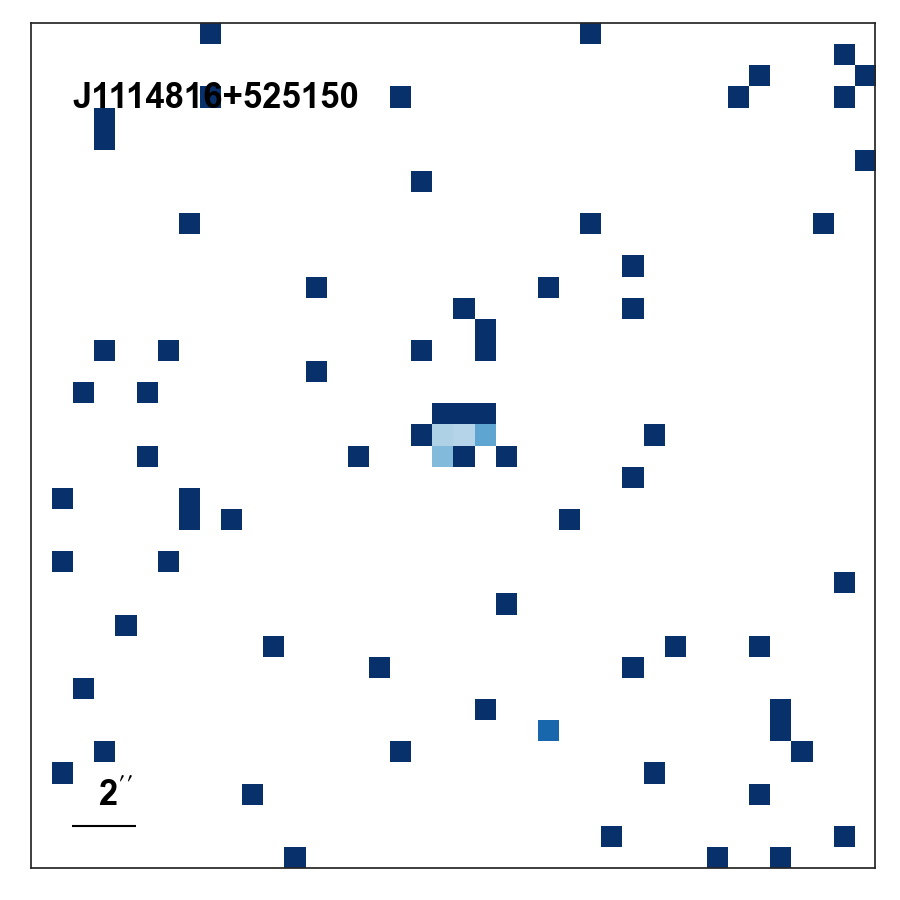}
     \end{subfigure}
     
    \vspace{0.2cm} 
     
    \begin{subfigure}
         \centering
         \includegraphics[width=0.3\textwidth]{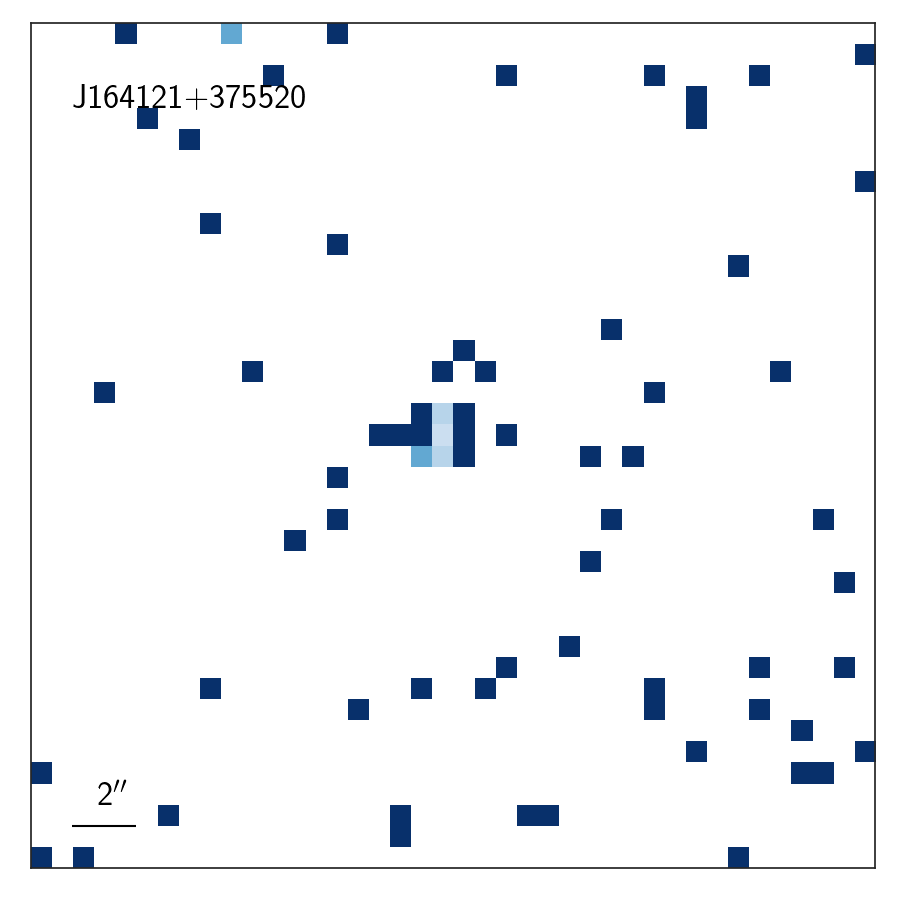}
    \end{subfigure}
    \hfill
    \begin{subfigure}
         \centering
         \includegraphics[width=0.3\textwidth]{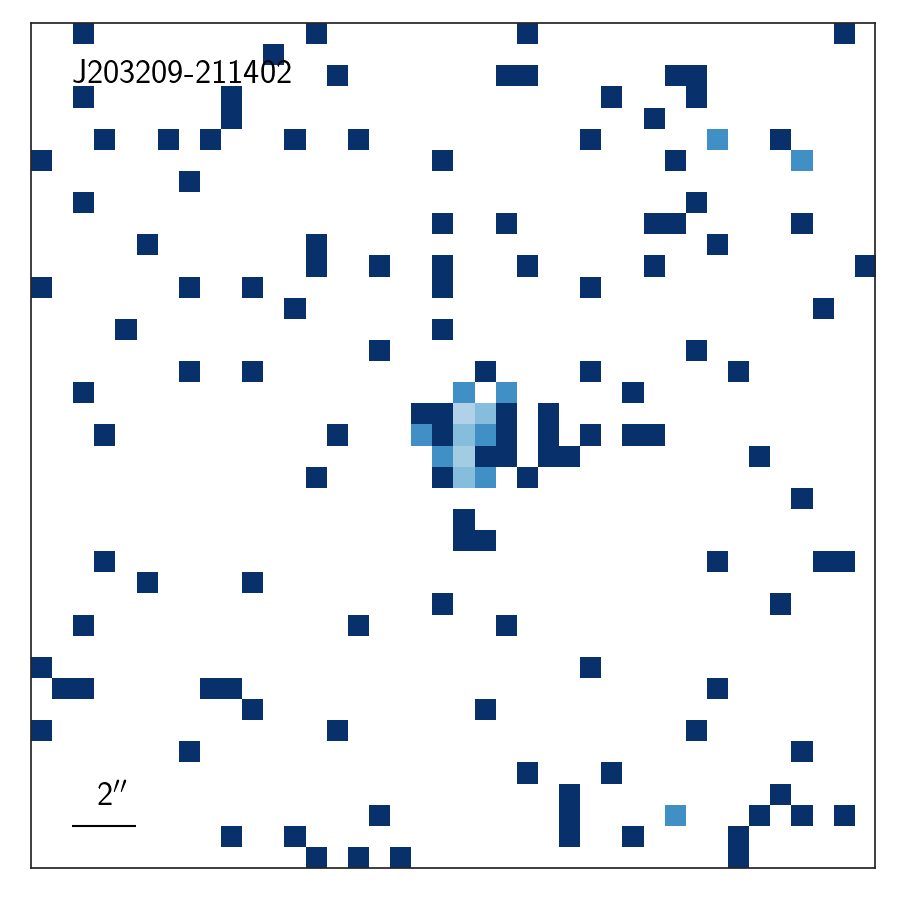}
    \end{subfigure}
    \hfill
    \begin{subfigure}
         \centering
         \includegraphics[width=0.3\textwidth]{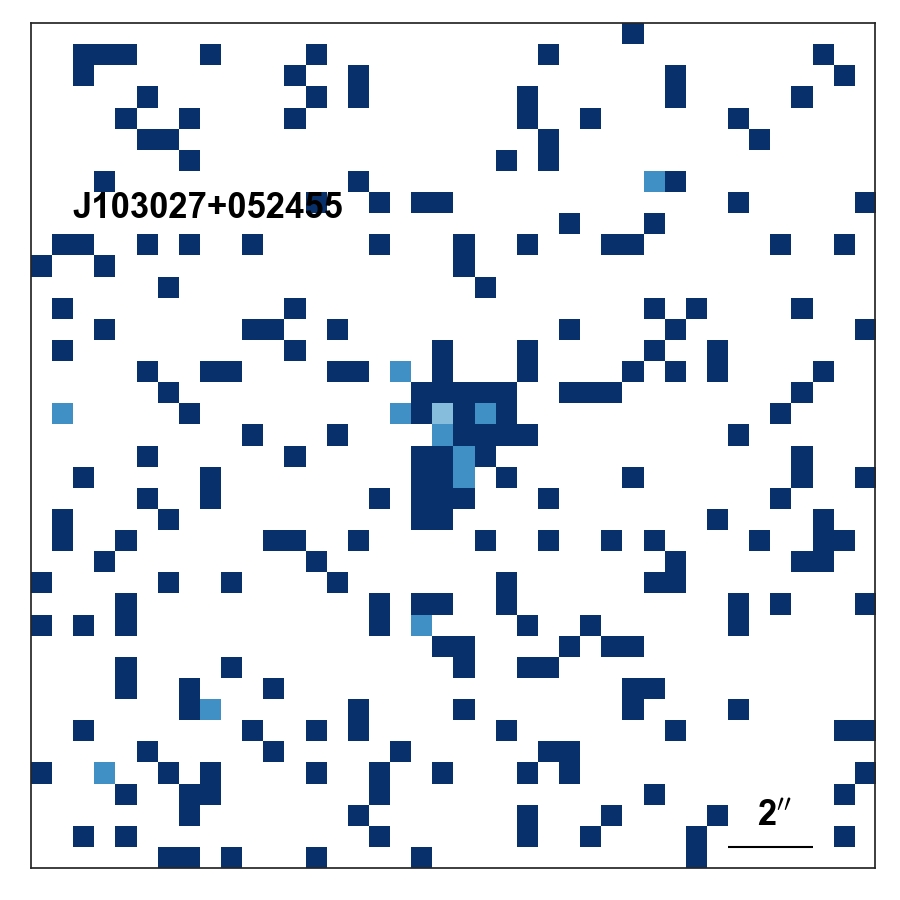}
     \end{subfigure}
     
      \vspace{0.2cm} 
     
    \begin{subfigure}
         \centering
         \includegraphics[width=0.3\textwidth]{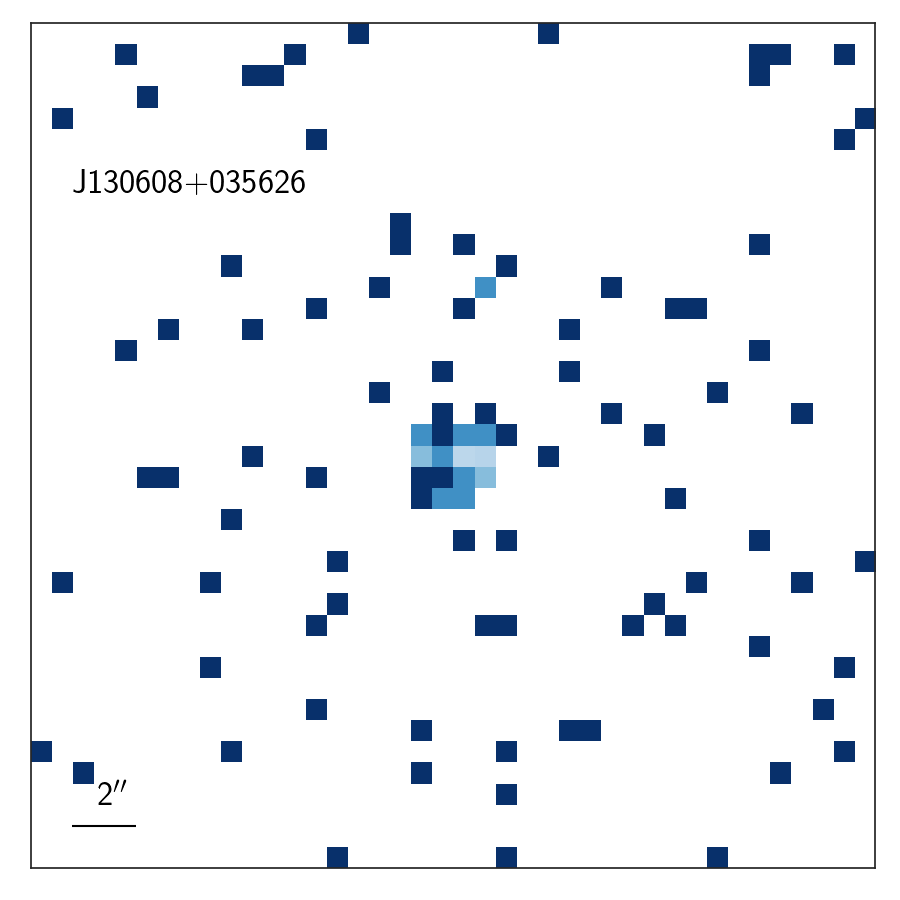}
    \end{subfigure}

        \caption{Binned X-ray images of selected sources in our dataset. Photons have energies in the range $0.5-8$ keV, and the scale is $0.5$\arcsec per pixel.}
        \label{fig:X_images2}
\end{figure*}

\begin{figure*}
\centering
     \begin{subfigure}
         \centering
         \includegraphics[width=0.45\textwidth]{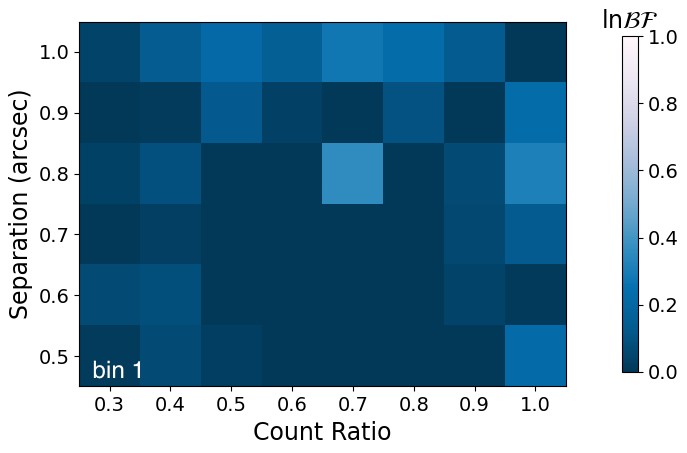}
     \end{subfigure}
     \hfill
     \begin{subfigure}
         \centering
         \includegraphics[width=0.45\textwidth]{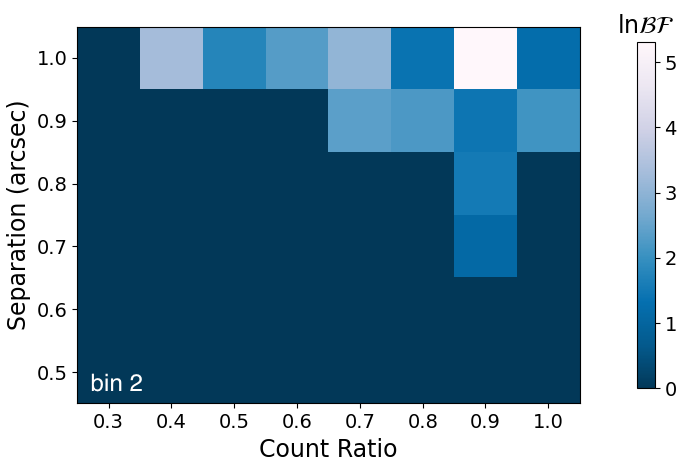}
     \end{subfigure}
    
     \vspace{-0.1cm}
     \begin{subfigure}
         \centering
         \includegraphics[width=0.45\textwidth]{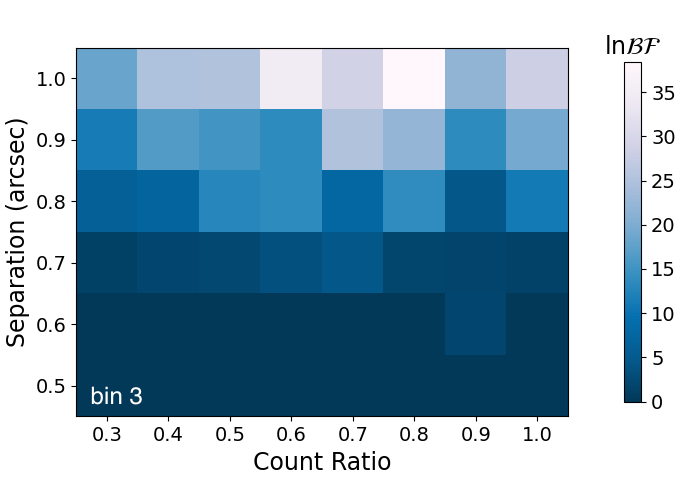}
     \end{subfigure}
     
     \vspace{0.2cm}
\caption{Results of the feasibility analysis described in Sec. \ref{subsec:sensitivity}. We show the sensitivty of \baymax to detect dual X-ray point sources in 3 count bins, as a function of separation ($r$) and count ratio ($f$). For our lowest count bin (average 9 counts), we are not sensitive to detecting dual X-ray point sources over most of parameter space. However, for bin 2 (average of 38 counts) at typical separations $r\ge1\arcsec$ we obtain Bayes factors that are compatible with a significant (at the 95\% confidence level, C.L.) detection; for bin 3 (average of 132 counts) at typical separations $r\ge0\farcs{7}$ we obtain Bayes factors that are compatible with a significant (at the 95\% C.L.) detection.} 
\label{fig:dual_sim_suite}
\end{figure*}

\section{RESULTS} 
\label{sec:results}
We are now in a position to discuss the results of our analysis performed with \baymax, and their consequences for the QLF.

\subsection{Results of the \baymax Analysis}
For each quasar, we restrict our analysis to photons with energies between 0.5$-$8 keV.  We analyze the photons contained within 20\arcsec $\times$ 20\arcsec~rectangular regions that are centered on the nominal X-ray coordinates of the quasar (see Figure~\ref{fig:X_images} and \ref{fig:X_images2}). We run \baymax using non-informative priors, where the prior distributions for location of the primary and secondary X-ray point source are uniform distributions bound in the x- and y-direction between the full extent of the 20\arcsec $\times$ 20\arcsec ~field of view. For quasars with multiple observations, \baymax models the PSF of each observation and calculates the likelihoods for each observation individually and then fits for astrometric shifts between different observations of the same source.

We ran \baymax 100 times on each of the 22 quasars reported in Table \ref{table:sources}, with the exception of J111033-132945, J152637-205000, J103027+052455, and J130608+035626. The reason to run the code a large number of times is that the choice of the parameters of the models is, by definition, stochastic. Hence, the computed Bayes factor is never exactly the same and an averaging process over a large number of runs is necessary to obtain accurate results, including an appropriate description of the confidence intervals. 

For J111033-132945, J152637-205000, J103027+052455, and J130608+035626, the computational time necessary to evaluate the sources 100 times was exceptionally large, a result of the total number of counts and/or number of observations per source. For these sources, we instead quote the statistical error bars returned from {\tt nestle} on a single-run. The errors provided by {\tt nestle} are calculated for each model, at each iteration, and are assumed to be proportional to the ratio of the ``information" to the number of live points (see \citealt{Skilling_2004} for explicit details). In the past, we have found that the statistical error bars returned from {\tt nestle} are consistent with the 1$\sigma$ spread in the $\ln{\mathcal{BF}}$ values when running \baymax 100 times on a single source \citep{Foord_2020, Foord_2021b}. 

In Table \ref{table:sources} we list the average $\ln{\mathcal{BF}}$ and the 1$\sigma$ error bar for each quasar. We find that all of the sources have $\ln{\mathcal{BF}}$ values consistent with 0 within the 1$\sigma$ error bar, indicative that \textit{there is no strong evidence for a lensed quasar in our sample}.
In the next \S \ref{subsec:sensitivity} we discuss the search sensitivity, which is crucial to correctly interpret this result.

\subsection{Search Sensitivity}
\label{subsec:sensitivity}
Considering that this paper presents the first use of \baymax to study lensed sources at very high-redshift and with low photon counts, we perform a full analysis of the sensitivity of this search.

For a given number of total counts, \baymax will be insensitive to multi-point sources at the low-end of separation and count ratio space. For example, for on-axis Chandra observations with $>700$ counts between 0.5$-$8 keV, we can expect that \baymax will be sensitive to dual point sources with count ratios greater than 0.2, down to separations as low as 0\farcs{3} \citep{Foord_2019}. However, at the low-end of the count range, \baymax may not be able to discern a single point source from multiple point sources across a wider range of count ratio and separation values. 

We measure the sensitivity of \baymax across our sample to quantify how well \baymax can correctly identify a putative lensed quasar, at a given separation and count ratio. Because the sensitivity of \baymax is directly tied to the total number of counts associated with a source, we divide our sample into 3 bins based on total number of 0.5$-$8 keV counts and analyze the sensitivity of \baymax for the average number of counts per bin. In Table~\ref{table:sources} we denote the sources in each bin via horizontal line breaks; bin 1 has fourteen sources with 3$-$15 counts, bin 2 has six sources with 21$-$72 counts, and bin 3 has two sources with $>$100 counts. 

We simulate dual X-ray point sources across $r$--$f$ space that have, on average, 9 (bin 1), 38 (bin 2), or 132 (bin 3) counts between 0.5$-$8 keV. We simulate systems with separations that range between 0\farcs{5}$-$1\arcsec and count ratios that range between 0.3$-$1. For each  $r$--$f$ point in the parameter space, we evaluate simulations with randomized position angles between the primary and secondary X-ray point source. Our results are shown in Figure~\ref{fig:dual_sim_suite}, where we plot the mean $\ln{\mathcal{BF}}$ for each point in the parameter space. 

For each count bin, we define regions of parameter space where \baymax cannot statistically differentiate between a single and multiple point source wherever the mean $\ln{\mathcal{BF}}$ is consistent with 0 at the 95\% confidence interval. This allows us to quantify regions in parameter space where each quasar is consistent with emission from a single X-ray point source. Consistent with expectations, \baymax favors the dual point source model more strongly as the separation and count ratio increase, and is capable of probing smaller separations for a given count ratio as the total number of counts increase. For bin 1, we find that at all count ratio and separation bins the mean $\ln{\mathcal{BF}}$ is consistent with 0. For bin 2, we find that at $r\ge1\arcsec$ we are sensitive to duals at $f\ge0.4$ and we can statistically probe separations as small as 0\farcs{7} at $f>0\farcs{7}$. Lastly, for bin 3, at $r\ge0\farcs{7}$ we are statistically sensitive to dual X-ray point sources across the entire $f$-space. 

Thus, for the majority of this sample (which reside in count bin 1), there is not enough data to claim at the 95\% confidence interval that the emission from the quasar is a consistent with single X-ray point source. However, at the 95\% C.L. we conclude that none of the quasars in bin 2 are dual at $r\ge1\arcsec$ for $f\ge0.4$ and none of the quasars in bin 3 are dual at $r\ge0\farcs{7}$ for $f>0.3$.

\begin{figure}
\includegraphics[angle=0,width=0.49\textwidth]{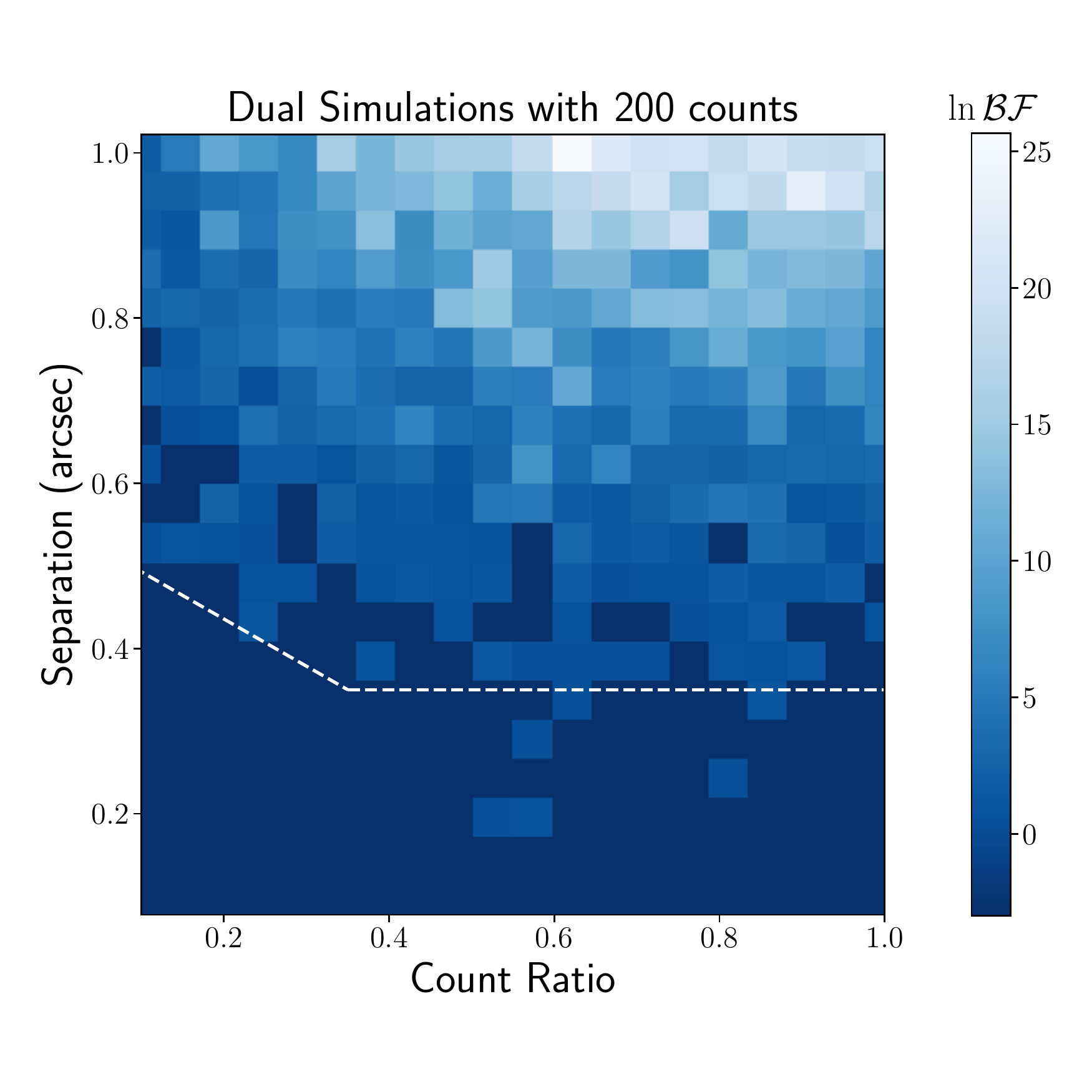}
\vspace{-1.25cm}
\caption{Bayes factors for simulated dual AGN with 200 total counts, with varying separation (in arcseconds) and count ratios. For each point in the parameter space, we evaluated 100 simulations with randomized position angles (0-360 $\deg$) between the primary and secondary image. Here we display the logarithm of the mean $\mathcal{BF}$ for each point in the parameter space. We enforce a cut of $\ln{\mathcal{BF}} > 3$, where above this value the Bayes factor is classified as strongly in favor of the dual point source model. Points in the parameter space with a $\ln{\mathcal{BF}}$ below this value are displayed in dark blue. With 200 counts between 0.5$-$8 keV, \baymax is capable of correctly identifying multiple AGN at flux ratios as low as $f=0.1$ down to separations of 0\farcs{5}. To reach this sensitivity for our sample, observations between 1.5 to 67 times deeper are necessary.}
\label{fig:dual_sim_suite_200counts}
\end{figure}

Carrying out simulations of dual X-ray point sources across $r$-$f$ space, while varying the total number of counts, we find that we are sensitive to detecting multiple X-ray point sources across most of $f$ space to separations as low as 0\farcs{5}, with at least 200 counts (see Figure~\ref{fig:dual_sim_suite_200counts}). To reach this sensitivity for each source, observations between 1.5$\times$ (at the low-end, for J130608+035626 and J103027+052455) to 67$\times$ (at the high-end, for J002429+391318) deeper are necessary, with the average observation needing to be 14$\times$ deeper. Given our average exposure time of 26 ks, such observations are not easily feasible with follow-up \emph{Chandra} observations. Alternatively, if each source in our sample was between 1.5$-$67 times brighter, the archival data sets would allow for a more sensitive study.

\subsection{Constraining the Shape of the Quasar Luminosity Function}
We now proceed to place some constraints on the value of the $\beta$ parameter for the QLF at $z\sim 6$. 

\begin{figure}
\includegraphics[angle=0,width=0.49\textwidth]{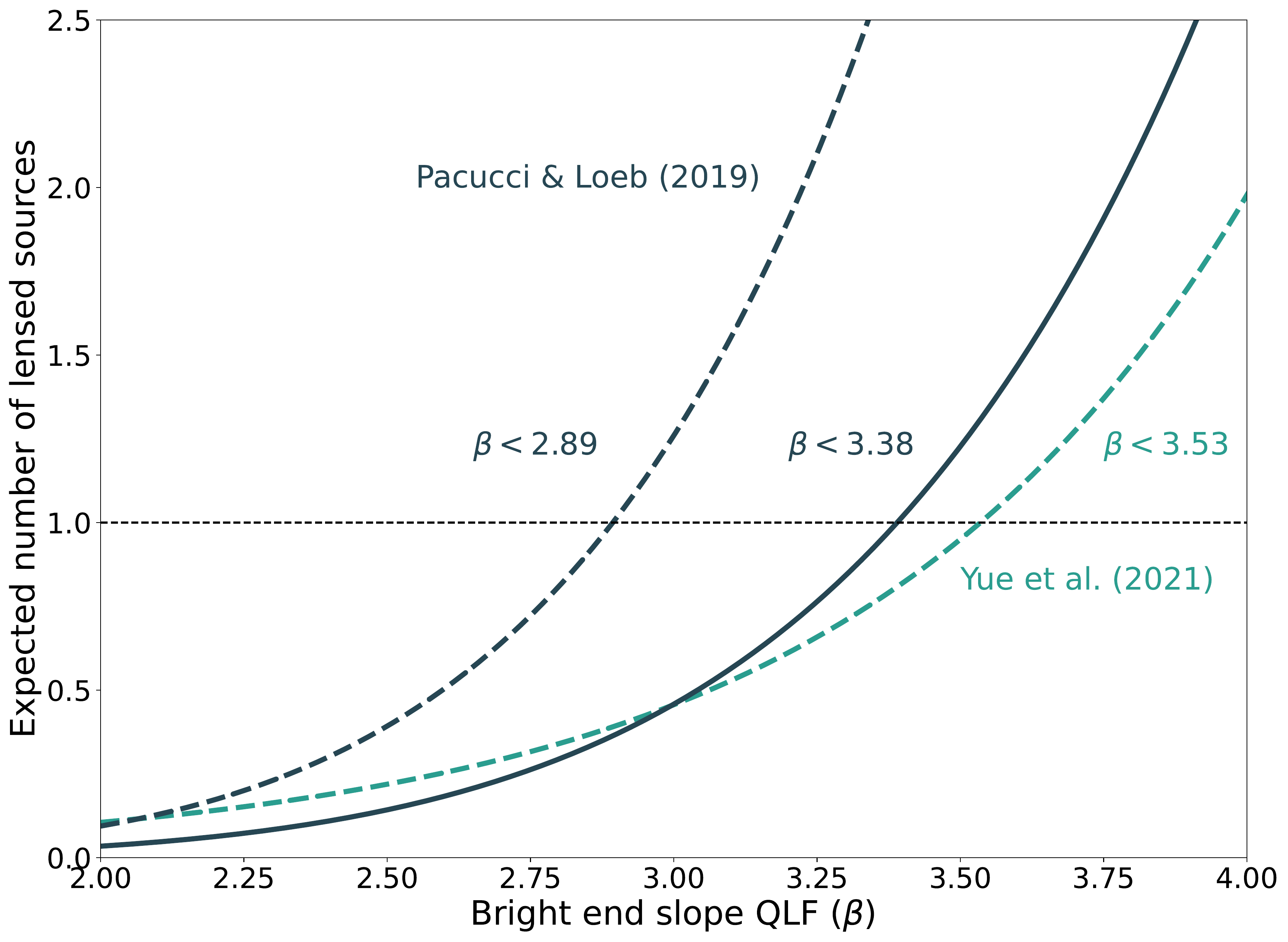}
\caption{Expected number of multiple sources in our sample, as a function of the bright-end slope, $\beta$, of the QLF. Predictions are based on the \high model (dark blue) and the \low model (green) as described in the main text. 
Considering only the 8 sources pertaining to the two bins with the highest count numbers, we obtain an upper limit of $\beta < 3.38$ in the \high model, while the \low model is not constraining.
With no detection of multiple sources out of the full sample of 22 sources (dashed lines), we obtain a constraint of $\beta < 2.89$ with the \high model, and $\beta < 3.53$ with the \low model.}
\label{fig:beta_constraint}
\end{figure}

As mentioned in \S \ref{sec:data}, \cite{Yue_2021} calculate that $\sim 85\%$ of sources at $z \sim 6$ should have a sufficient separation between images to be detected above the resolution limit $\varphi_{\rm lim} \sim 0.\arcsec 5$ of the Chandra X-ray telescope (see a similar calculation at $z\approx 6.5$ also in \citealt{Pacucci_Loeb_2019}). We call this probability $P(\varphi > \varphi_{\rm lim})$.
Additionally, for each value of $\beta$ in the range $[2,4]$, we can use the \high model and the \low model to compute the lensed fraction $P(\beta)$. 
Hence, the fraction of N sources with sufficient angular separation to be observed as multiples by Chandra, $N \times P(\varphi > \varphi_{\rm lim})$, is multiplied by the probability $P(\beta)$, for values of $\beta$ in the range $[2,4]$, thus obtaining Fig. \ref{fig:beta_constraint}.

In this figure we show with a solid line the expected number of multiple sources in the sample of $N=8$ Chandra-detected quasars with the highest photon counts, as a function of the parameter $\beta$. The sources with the highest photon number counts, reported in Table \ref{table:sources}, are the only ones for which we can exclude multiplicity, following our analysis with \baymax. 
As we cannot claim any multiple source with high significance in this reduced dataset, the number of observed lensed quasars is $N_{\rm obs} < 1$. This model leads to an upper limit of $\beta < 3.38$, which excludes at $95\%$ C.L. the value of $\beta = 3.6$ \citep{Yang_2016} commonly used in the literature.
Unfortunately, only 8 sources are not sufficient to place any meaningful upper limit with the \low model.

Considering the full sample of $N=22$ sources, we can also calculate constraints on $\beta$, in the assumption that none of them are multiple. This translates in the following \textit{upper limits} on $\beta$:
\begin{itemize}
    \item $\beta < 2.89$ in the \high model
    \item $\beta < 3.53$ in the \low model
\end{itemize}

The stricter upper limit rules out the commonly used value of $\beta = 3.6$ \citep{Yang_2016}, while it is compatible with the shallower value of $\beta = 2.8$ \citep{Jiang_2016}. The looser upper limit also excludes $\beta = 3.6$. Finally, note that overall our study significantly rules out steeper values of the bright-end slope of the QLF, e.g., $\beta \approx 5$ \citep{Kulkarni_2018}.

Assuming $\beta\approx2.8$, a value that is widely used in the literature (e.g., \citealt{Jiang_2016}) and in accordance with our limits, the probability of a $z\sim$ 6 quasar being lensed is estimated to be approximately 2\% and 4\% by the \low and \high models, respectively. Assuming that all our observations meet the count threshold ($>$200 counts between 0.5$-$8 keV) necessary for a sensitivity down to separations of 0\farcs{5}, we estimate that we would need a sample size between 14$-$27 high-$z$ AGN to have $>$10\% chance of detecting 1 or more lensed AGN (assuming binomial statistics). The sample size increases to be between 42$-$84 high-$z$ AGN to have $>$50\% chance of detecting 1 or more lensed AGN. Given the large exposure times necessary to detect $>$200 counts for high-$z$ AGN with \emph{Chandra}, such a project is likely not currently feasible. However, future X-ray telescopes with similar (or better) angular resolution to \emph{Chandra}, but larger sensitivity, may allow for such an observing project. The probe-class mission concept Advanced X-ray Imaging Satellite (AXIS), for example, will have a collecting area about an order of magnitude larger than Chandra \citep{AXIS_2018, AXIS_dual_2019}.

\section{Discussion and Conclusions} 
\label{sec:disc_concl}
In this study we have investigated 22 X-ray detected quasars at $z >5.8$ with \baymax, a code designed to perform a Bayesian analysis of multiple sources, whether the multiple images are physical or caused by gravitational lensing. Our goal was to study the lensed fraction of high-$z$ quasars, after the discovery of the first strongly lensed quasar by \citealt{Fan_2019}. 

\textit{Out of 22 sources, we could not confirm any statistically significant multiple image.} For the 8 sources with a photon count number $>20$, we can rule out at $95\%$ C.L. that they have multiple images with separations larger than $1\arcsec$ and count ratios $f>0.4$.
While some of the sources analyzed were flagged by \baymax as possessing a slight preference for the multiple image model, all of them were statistically compatible with the single model.

This non-detection allows us to constrain the bright-end slope of the QLF at $z \sim 6$. With only the $N=8$ sources pertaining in the two highest count bins, we obtain an upper limit of $\beta < 3.38$ with the \high model. Considering $N=22$ sources, we obtain $\beta < 2.89$ in the \high model and $\beta < 3.53$ in the \low model.
These upper limits significantly rule out extreme values of $\beta$ recently proposed in the literature (e.g., $\beta \approx 5$, \citealt{Kulkarni_2018}), and point to much shallower values (e.g., $\beta = 2.8$, \citealt{Jiang_2016}).

Is it surprising that we found no clear sign of lensed quasars in our sample of $z \gtrsim 6$ sources? Possibly not. Some of the possible reasons to explain why we found none are the following:
\begin{enumerate}
    \item Higher photon counts are needed, requiring deeper Chandra observations, next-generation X-ray telescopes, or brighter sources.
    \item Larger source samples are needed.
    \item The lensed fraction of high-$z$ quasars is lower than previously predicted \citep{Yue_2021}.
\end{enumerate}

Regarding (i), our search sensitivity analysis with \baymax (see Sec. \ref{subsec:sensitivity}) showed that we need photon counts $>20$ to be able to discriminate sources with separations $r > 1\arcsec$ and count ratios $f>0.4$ at $95\%$ C.L.; unfortunately, only $\sim 36\%$ of the sources in our sample met this criterion. Deeper Chandra observations of known quasars, or a sample selection of only the X-ray brightest sources, can improve the statistics for future studies of this kind.

Additionally, we chose a uniform sample of $22$ Chandra-detected sources which are on-axis and some of them with multiple observations. By expanding the sample significantly, we might have ended up with a statistically significant detection. For example, \cite{Yue_2021} predicts that there should be $\sim 15$ gravitationally lensed quasars at $z > 6$ currently detectable (in optical wavelengths, at least) in the whole sky.

Ultimately, it is possible that previous theoretical models (e.g, \citealt{Wyithe_Loeb_2002, Comerford_2002, Pacucci_Loeb_2019}) over-estimated the lensed fraction of quasars at $z \sim 6$. Recently, \cite{Yue_2021} suggested that the lensed fraction could be $\sim 10$ times lower than previously predicted, explaining the observational detection of a single lensed quasar in the epoch of reionization thus far \citep{Fan_2019}.
This would suggest the necessity to re-evaluate the velocity dispersion function of galaxies at $z \lesssim 3$, i.e. where most of the lensing sources are. In fact, \cite{Yue_2021} show that the observed low number of lensed quasars at $z>6$ can be explained by using the locally observed velocity dispersion function at $z < 0.1$ \citep{Sohn_2017, Hasan_2019}. Previous studies may have thus over-estimated the available lensing power at $z \lesssim 3$.

Upcoming optical surveys such as Euclid and the LSST will greatly expand our reach in the study of lensed quasars at $z > 6$, and will most likely solve the long-lasting controversy over the lensed fraction. The expansion of our search of lensed quasars to optically-observed sources would require a substantial re-purposing of \baymax, to include statistics able to deal with very high photon counts. The possibilities opened up by a Bayesian analysis of a quasar image, when compared to a visual inspection, are worth investigating in the near future.
Whatever route we take, it is most likely that with upcoming optical and X-ray surveys of the sky the multi-decadal question of how likely it is for a high-$z$ quasar to be strongly lensed will soon be solved.

\section*{Acknowledgements}
We thank the anonymous referee for constructive comments on the manuscript.
F.P. acknowledges support from a Clay Fellowship administered by the Smithsonian Astrophysical Observatory.
This work was supported by the Black Hole Initiative at Harvard University, which is funded by grants from the John Templeton Foundation and the Gordon and Betty Moore Foundation.
A.F. acknowledges support by the Porat Postdoctoral Fellowship at Stanford University.

\section*{Data Availability}
The code \href{https://www.adifoord.com/research/agn-pairs}{\baymax} is proprietary. The scripts used to analyze the data will be shared on reasonable request to the corresponding authors.



\bibliographystyle{mnras}
\bibliography{ms} 

\begin{thebibliography}{}
\makeatletter
\relax
\def\mn@urlcharsother{\let\do\@makeother \do\$\do\&\do\#\do\^\do\_\do\%\do\~}
\def\mn@doi{\begingroup\mn@urlcharsother \@ifnextchar [ {\mn@doi@}
  {\mn@doi@[]}}
\def\mn@doi@[#1]#2{\def\@tempa{#1}\ifx\@tempa\@empty \href
  {http://dx.doi.org/#2} {doi:#2}\else \href {http://dx.doi.org/#2} {#1}\fi
  \endgroup}
\def\mn@eprint#1#2{\mn@eprint@#1:#2::\@nil}
\def\mn@eprint@arXiv#1{\href {http://arxiv.org/abs/#1} {{\tt arXiv:#1}}}
\def\mn@eprint@dblp#1{\href {http://dblp.uni-trier.de/rec/bibtex/#1.xml}
  {dblp:#1}}
\def\mn@eprint@#1:#2:#3:#4\@nil{\def\@tempa {#1}\def\@tempb {#2}\def\@tempc
  {#3}\ifx \@tempc \@empty \let \@tempc \@tempb \let \@tempb \@tempa \fi \ifx
  \@tempb \@empty \def\@tempb {arXiv}\fi \@ifundefined
  {mn@eprint@\@tempb}{\@tempb:\@tempc}{\expandafter \expandafter \csname
  mn@eprint@\@tempb\endcsname \expandafter{\@tempc}}}

\bibitem[\protect\citeauthoryear{{Alcock} et~al.,}{{Alcock}
  et~al.}{1996}]{Alcock_1996}
{Alcock} C.,  et~al., 1996, \mn@doi [\apj] {10.1086/178005}, \href
  {https://ui.adsabs.harvard.edu/abs/1996ApJ...471..774A} {471, 774}

\bibitem[\protect\citeauthoryear{{Ba{\~n}ados} et~al.,}{{Ba{\~n}ados}
  et~al.}{2016}]{Banados_2016}
{Ba{\~n}ados} E.,  et~al., 2016, \mn@doi [\apjs] {10.3847/0067-0049/227/1/11},
  \href {https://ui.adsabs.harvard.edu/abs/2016ApJS..227...11B} {227, 11}

\bibitem[\protect\citeauthoryear{{Comerford}, {Haiman}  \&
  {Schaye}}{{Comerford} et~al.}{2002}]{Comerford_2002}
{Comerford} J.~M.,  {Haiman} Z.,   {Schaye} J.,  2002, \mn@doi [\apj]
  {10.1086/343116}, \href
  {https://ui.adsabs.harvard.edu/#abs/2002ApJ...580...63C} {580, 63}

\bibitem[\protect\citeauthoryear{{Davies}, {Wang}, {Eilers}  \&
  {Hennawi}}{{Davies} et~al.}{2020}]{Davies_2020}
{Davies} F.~B.,  {Wang} F.,  {Eilers} A.-C.,   {Hennawi} J.~F.,  2020, \mn@doi
  [\apjl] {10.3847/2041-8213/abc61f}, \href
  {https://ui.adsabs.harvard.edu/abs/2020ApJ...904L..32D} {904, L32}

\bibitem[\protect\citeauthoryear{{Fan} et~al.}{{Fan} et~al.}{2001}]{Fan_2001}
{Fan} X.,  et~al., 2001, \mn@doi [\aj] {10.1086/324111}, \href
  {http://adsabs.harvard.edu/abs/2001AJ....122.2833F} {122, 2833}

\bibitem[\protect\citeauthoryear{{Fan} et~al.,}{{Fan} et~al.}{2003}]{Fan_2003}
{Fan} X.,  et~al., 2003, \mn@doi [\aj] {10.1086/368246}, \href
  {https://ui.adsabs.harvard.edu/abs/2003AJ....125.1649F} {125, 1649}

\bibitem[\protect\citeauthoryear{{Fan} et~al.,}{{Fan} et~al.}{2004}]{Fan_2004}
{Fan} X.,  et~al., 2004, \mn@doi [\aj] {10.1086/422434}, \href
  {https://ui.adsabs.harvard.edu/abs/2004AJ....128..515F} {128, 515}

\bibitem[\protect\citeauthoryear{{Fan} et~al.}{{Fan} et~al.}{2006}]{Fan_2006}
{Fan} X.,  et~al., 2006, \mn@doi [\aj] {10.1086/500296}, \href
  {http://adsabs.harvard.edu/abs/2006AJ....131.1203F} {131, 1203}

\bibitem[\protect\citeauthoryear{{Fan} et~al.,}{{Fan} et~al.}{2019}]{Fan_2019}
{Fan} X.,  et~al., 2019, \mn@doi [\apjl] {10.3847/2041-8213/aaeffe}, \href
  {https://ui.adsabs.harvard.edu/abs/2019ApJ...870L..11F} {870, L11}

\bibitem[\protect\citeauthoryear{{Foord} et~al.,}{{Foord}
  et~al.}{2019}]{Foord_2019}
{Foord} A.,  et~al., 2019, \mn@doi [\apj] {10.3847/1538-4357/ab18a3}, \href
  {https://ui.adsabs.harvard.edu/abs/2019ApJ...877...17F} {877, 17}

\bibitem[\protect\citeauthoryear{{Foord}, {G{\"u}ltekin}, {Nevin}, {Comerford},
  {Hodges-Kluck}, {Barrows}, {Goulding}  \& {Greene}}{{Foord}
  et~al.}{2020}]{Foord_2020}
{Foord} A.,  {G{\"u}ltekin} K.,  {Nevin} R.,  {Comerford} J.~M.,
  {Hodges-Kluck} E.,  {Barrows} R.~S.,  {Goulding} A.~D.,   {Greene} J.~E.,
  2020, \mn@doi [\apj] {10.3847/1538-4357/ab72fa}, \href
  {https://ui.adsabs.harvard.edu/abs/2020ApJ...892...29F} {892, 29}

\bibitem[\protect\citeauthoryear{{Foord}, {G{\"u}ltekin}, {Runnoe}  \&
  {Koss}}{{Foord} et~al.}{2021}]{Foord_2021b}
{Foord} A.,  {G{\"u}ltekin} K.,  {Runnoe} J.~C.,   {Koss} M.~J.,  2021, \mn@doi
  [\apj] {10.3847/1538-4357/abce5d}, \href
  {https://ui.adsabs.harvard.edu/abs/2021ApJ...907...71F} {907, 71}

\bibitem[\protect\citeauthoryear{{Fujimoto}, {Oguri}, {Nagao}, {Izumi}  \&
  {Ouchi}}{{Fujimoto} et~al.}{2020}]{Fujimoto_2020}
{Fujimoto} S.,  {Oguri} M.,  {Nagao} T.,  {Izumi} T.,   {Ouchi} M.,  2020,
  \mn@doi [\apj] {10.3847/1538-4357/ab718c}, \href
  {https://ui.adsabs.harvard.edu/abs/2020ApJ...891...64F} {891, 64}

\bibitem[\protect\citeauthoryear{{Gallerani}, {Fan}, {Maiolino}  \&
  {Pacucci}}{{Gallerani} et~al.}{2017}]{Gallerani_2017}
{Gallerani} S.,  {Fan} X.,  {Maiolino} R.,   {Pacucci} F.,  2017, \mn@doi
  [\pasa] {10.1017/pasa.2017.14}, \href
  {https://ui.adsabs.harvard.edu/abs/2017PASA...34...22G} {34, e022}

\bibitem[\protect\citeauthoryear{{Gunn} \& {Peterson}}{{Gunn} \&
  {Peterson}}{1965}]{Gunn_Peterson_1965}
{Gunn} J.~E.,  {Peterson} B.~A.,  1965, \mn@doi [\apj] {10.1086/148444}, \href
  {https://ui.adsabs.harvard.edu/abs/1965ApJ...142.1633G} {142, 1633}

\bibitem[\protect\citeauthoryear{{Haiman}}{{Haiman}}{2013}]{Haiman_2013}
{Haiman} Z.,  2013, in {Wiklind} T.,  {Mobasher} B.,   {Bromm} V.,  eds,
  Astrophysics and Space Science Library Vol. 396, The First Galaxies. p.~293
  (\mn@eprint {arXiv} {1203.6075}), \mn@doi{10.1007/978-3-642-32362-1\_6}

\bibitem[\protect\citeauthoryear{{Haiman} \& {Quataert}}{{Haiman} \&
  {Quataert}}{2004}]{Haiman_Quataert_2004}
{Haiman} Z.,  {Quataert} E.,  2004, in {Barger} A.~J.,  ed.,  Astrophysics and
  Space Science Library Vol. 308, Supermassive Black Holes in the Distant
  Universe. p.~147 (\mn@eprint {arXiv} {astro-ph/0403225}),
  \mn@doi{10.1007/978-1-4020-2471-9\_5}

\bibitem[\protect\citeauthoryear{{Hasan} \& {Crocker}}{{Hasan} \&
  {Crocker}}{2019}]{Hasan_2019}
{Hasan} F.,  {Crocker} A.,  2019, arXiv e-prints, \href
  {https://ui.adsabs.harvard.edu/abs/2019arXiv190400486H} {p. arXiv:1904.00486}

\bibitem[\protect\citeauthoryear{{Hilbert}, {White}, {Hartlap}  \&
  {Schneider}}{{Hilbert} et~al.}{2007}]{Hilbert_2007}
{Hilbert} S.,  {White} S. D.~M.,  {Hartlap} J.,   {Schneider} P.,  2007,
  \mn@doi [\mnras] {10.1111/j.1365-2966.2007.12391.x}, \href
  {https://ui.adsabs.harvard.edu/abs/2007MNRAS.382..121H} {382, 121}

\bibitem[\protect\citeauthoryear{{Inayoshi}, {Visbal}  \& {Haiman}}{{Inayoshi}
  et~al.}{2020}]{Inayoshi_review_2019}
{Inayoshi} K.,  {Visbal} E.,   {Haiman} Z.,  2020, \mn@doi [\araa]
  {10.1146/annurev-astro-120419-014455}, \href
  {https://ui.adsabs.harvard.edu/abs/2020ARA&A..58...27I} {58, 27}

\bibitem[\protect\citeauthoryear{{Jeffreys}}{{Jeffreys}}{1935}]{Jeffreys_1935}
{Jeffreys} H.,  1935, \mn@doi [Proceedings of the Cambridge Philosophical
  Society] {10.1017/S030500410001330X}, \href
  {https://ui.adsabs.harvard.edu/abs/1935PCPS...31..203J} {31, 203}

\bibitem[\protect\citeauthoryear{{Jiang} et~al.,}{{Jiang}
  et~al.}{2016}]{Jiang_2016}
{Jiang} L.,  et~al., 2016, \mn@doi [\apj] {10.3847/1538-4357/833/2/222}, \href
  {https://ui.adsabs.harvard.edu/#abs/2016ApJ...833..222J} {833, 222}

\bibitem[\protect\citeauthoryear{{Keeton}, {Kuhlen}  \& {Haiman}}{{Keeton}
  et~al.}{2005}]{Keeton_2005}
{Keeton} C.~R.,  {Kuhlen} M.,   {Haiman} Z.,  2005, \mn@doi [\apj]
  {10.1086/427722}, \href
  {https://ui.adsabs.harvard.edu/abs/2005ApJ...621..559K} {621, 559}

\bibitem[\protect\citeauthoryear{{Koss} et~al.,}{{Koss}
  et~al.}{2019}]{AXIS_dual_2019}
{Koss} M.,  et~al., 2019, Astro2020: Decadal Survey on Astronomy and
  Astrophysics, \href {https://ui.adsabs.harvard.edu/abs/2019astro2020T.504K}
  {2020, 504}

\bibitem[\protect\citeauthoryear{{Kulkarni}, {Worseck}  \&
  {Hennawi}}{{Kulkarni} et~al.}{2019}]{Kulkarni_2018}
{Kulkarni} G.,  {Worseck} G.,   {Hennawi} J.~F.,  2019, \mn@doi [\mnras]
  {10.1093/mnras/stz1493}, \href
  {https://ui.adsabs.harvard.edu/abs/2019MNRAS.488.1035K} {488, 1035}

\bibitem[\protect\citeauthoryear{{Li}, {Wang}, {Yang}, {Bregman}, {Fan}  \&
  {Zhang}}{{Li} et~al.}{2021}]{Li_2021}
{Li} J.-T.,  {Wang} F.,  {Yang} J.,  {Bregman} J.~N.,  {Fan} X.,   {Zhang} Y.,
  2021, \mn@doi [\mnras] {10.1093/mnras/stab1042}, \href
  {https://ui.adsabs.harvard.edu/abs/2021MNRAS.504.2767L} {504, 2767}

\bibitem[\protect\citeauthoryear{{Mazzucchelli} et~al.,}{{Mazzucchelli}
  et~al.}{2017}]{Mazzucchelli_2017}
{Mazzucchelli} C.,  et~al., 2017, \mn@doi [\apj] {10.3847/1538-4357/aa9185},
  \href {https://ui.adsabs.harvard.edu/abs/2017ApJ...849...91M} {849, 91}

\bibitem[\protect\citeauthoryear{{Meneghetti} et~al.,}{{Meneghetti}
  et~al.}{2020}]{Meneghetti_2020}
{Meneghetti} M.,  et~al., 2020, \mn@doi [Science] {10.1126/science.aax5164},
  \href {https://ui.adsabs.harvard.edu/abs/2020Sci...369.1347M} {369, 1347}

\bibitem[\protect\citeauthoryear{{Mortlock} et~al.,}{{Mortlock}
  et~al.}{2011}]{Mortlock_2011}
{Mortlock} D.~J.,  et~al., 2011, \mn@doi [\nat] {10.1038/nature10159}, \href
  {http://adsabs.harvard.edu/abs/2011Natur.474..616M} {474, 616}

\bibitem[\protect\citeauthoryear{{Mushotzky}}{{Mushotzky}}{2018}]{AXIS_2018}
{Mushotzky} R.,  2018, in {den Herder} J.-W.~A.,  {Nikzad} S.,   {Nakazawa} K.,
   eds,  Society of Photo-Optical Instrumentation Engineers (SPIE) Conference
  Series Vol. 10699, Space Telescopes and Instrumentation 2018: Ultraviolet to
  Gamma Ray. p. 1069929 (\mn@eprint {arXiv} {1807.02122}),
  \mn@doi{10.1117/12.2310003}

\bibitem[\protect\citeauthoryear{{Natarajan} et~al.,}{{Natarajan}
  et~al.}{2017}]{Natarajan_2017_lensing}
{Natarajan} P.,  et~al., 2017, \mn@doi [\mnras] {10.1093/mnras/stw3385}, \href
  {https://ui.adsabs.harvard.edu/abs/2017MNRAS.468.1962N} {468, 1962}

\bibitem[\protect\citeauthoryear{{Pacucci} \& {Loeb}}{{Pacucci} \&
  {Loeb}}{2019}]{Pacucci_Loeb_2019}
{Pacucci} F.,  {Loeb} A.,  2019, \mn@doi [\apjl] {10.3847/2041-8213/aaf86a},
  \href {https://ui.adsabs.harvard.edu/abs/2019ApJ...870L..12P} {870, L12}

\bibitem[\protect\citeauthoryear{{Pacucci} \& {Loeb}}{{Pacucci} \&
  {Loeb}}{2020}]{Pacucci_Loeb_2019_mirage}
{Pacucci} F.,  {Loeb} A.,  2020, \mn@doi [\apj] {10.3847/1538-4357/ab6130},
  \href {https://ui.adsabs.harvard.edu/abs/2020ApJ...889...52P} {889, 52}

\bibitem[\protect\citeauthoryear{{Pacucci} \& {Loeb}}{{Pacucci} \&
  {Loeb}}{2022}]{Pacucci_2022}
{Pacucci} F.,  {Loeb} A.,  2022, \mn@doi [\mnras] {10.1093/mnras/stab3071},
  \href {https://ui.adsabs.harvard.edu/abs/2022MNRAS.509.1885P} {509, 1885}

\bibitem[\protect\citeauthoryear{{Pei}}{{Pei}}{1993}]{Pei_1993}
{Pei} Y.~C.,  1993, \mn@doi [\apj] {10.1086/172176}, \href
  {https://ui.adsabs.harvard.edu/#abs/1993ApJ...403....7P} {403, 7}

\bibitem[\protect\citeauthoryear{{Pei}}{{Pei}}{1995}]{Pei_1995}
{Pei} Y.~C.,  1995, \mn@doi [\apj] {10.1086/175290}, \href
  {http://adsabs.harvard.edu/abs/1995ApJ...440..485P} {440, 485}

\bibitem[\protect\citeauthoryear{{Salvatier}, {Wiecki}  \&
  {Fonnesbeck}}{{Salvatier} et~al.}{2016}]{Salvatier_2016}
{Salvatier} J.,  {Wiecki} T.,   {Fonnesbeck} C.,  2016, \mn@doi [PeerJ Computer
  Science] {10.7717/peerj-cs.55.}, 2

\bibitem[\protect\citeauthoryear{{Schmidt}}{{Schmidt}}{1968}]{Schmidt_1968}
{Schmidt} M.,  1968, \mn@doi [\apj] {10.1086/149446}, \href
  {https://ui.adsabs.harvard.edu/abs/1968ApJ...151..393S} {151, 393}

\bibitem[\protect\citeauthoryear{{Schneider}, {Ehlers}  \& {Falco}}{{Schneider}
  et~al.}{1992}]{Schneider_1992}
{Schneider} P.,  {Ehlers} J.,   {Falco} E.~E.,  1992, {Gravitational Lenses},
  \mn@doi{10.1007/978-3-662-03758-4.
}

\bibitem[\protect\citeauthoryear{{Shen} et~al.,}{{Shen}
  et~al.}{2011}]{Shen_2011}
{Shen} Y.,  et~al., 2011, \mn@doi [\apjs] {10.1088/0067-0049/194/2/45}, \href
  {https://ui.adsabs.harvard.edu/abs/2011ApJS..194...45S} {194, 45}

\bibitem[\protect\citeauthoryear{{Skilling}}{{Skilling}}{2004}]{Skilling_2004}
{Skilling} J.,  2004, in {Fischer} R.,  {Preuss} R.,   {Toussaint} U.~V.,  eds,
   American Institute of Physics Conference Series Vol. 735, American Institute
  of Physics Conference Series. pp 395--405, \mn@doi{10.1063/1.1835238}

\bibitem[\protect\citeauthoryear{{Sohn}, {Zahid}  \& {Geller}}{{Sohn}
  et~al.}{2017}]{Sohn_2017}
{Sohn} J.,  {Zahid} H.~J.,   {Geller} M.~J.,  2017, \mn@doi [\apj]
  {10.3847/1538-4357/aa7de3}, \href
  {https://ui.adsabs.harvard.edu/abs/2017ApJ...845...73S} {845, 73}

\bibitem[\protect\citeauthoryear{{Tang} et~al.,}{{Tang}
  et~al.}{2017}]{Tang_2017}
{Tang} J.-J.,  et~al., 2017, \mn@doi [\mnras] {10.1093/mnras/stw3287}, \href
  {https://ui.adsabs.harvard.edu/abs/2017MNRAS.466.4568T} {466, 4568}

\bibitem[\protect\citeauthoryear{{Treu}}{{Treu}}{2010}]{Treu_2010}
{Treu} T.,  2010, \mn@doi [\araa] {10.1146/annurev-astro-081309-130924}, \href
  {https://ui.adsabs.harvard.edu/abs/2010ARA&A..48...87T} {48, 87}

\bibitem[\protect\citeauthoryear{{Turner}}{{Turner}}{1980}]{Turner_1980}
{Turner} E.~L.,  1980, \mn@doi [\apjl] {10.1086/183418}, \href
  {http://adsabs.harvard.edu/abs/1980ApJ...242L.135T} {242, L135}

\bibitem[\protect\citeauthoryear{{Venemans} et~al.,}{{Venemans}
  et~al.}{2015}]{Venemans_2015}
{Venemans} B.~P.,  et~al., 2015, \mn@doi [\apjl] {10.1088/2041-8205/801/1/L11},
  \href {https://ui.adsabs.harvard.edu/abs/2015ApJ...801L..11V} {801, L11}

\bibitem[\protect\citeauthoryear{{Volonteri}}{{Volonteri}}{2010}]{Volonteri_2010}
{Volonteri} M.,  2010, \mn@doi [{Astronomy and Astrophysics Review}]
  {10.1007/s00159-010-0029-x}, \href
  {http://adsabs.harvard.edu/abs/2010A%26ARv..18..279V} {18, 279}

\bibitem[\protect\citeauthoryear{{Wang} et~al.,}{{Wang}
  et~al.}{2021}]{Wang_2021}
{Wang} F.,  et~al., 2021, \mn@doi [\apjl] {10.3847/2041-8213/abd8c6}, \href
  {https://ui.adsabs.harvard.edu/abs/2021ApJ...907L...1W} {907, L1}

\bibitem[\protect\citeauthoryear{{Willott} et~al.,}{{Willott}
  et~al.}{2007}]{Willott_2007}
{Willott} C.~J.,  et~al., 2007, \mn@doi [\aj] {10.1086/522962}, \href
  {https://ui.adsabs.harvard.edu/abs/2007AJ....134.2435W} {134, 2435}

\bibitem[\protect\citeauthoryear{{Willott} et~al.,}{{Willott}
  et~al.}{2010}]{Willott_2010}
{Willott} C.~J.,  et~al., 2010, \mn@doi [\aj] {10.1088/0004-6256/139/3/906},
  \href {https://ui.adsabs.harvard.edu/abs/2010AJ....139..906W} {139, 906}

\bibitem[\protect\citeauthoryear{{Wong} et~al.,}{{Wong}
  et~al.}{2020}]{holicow_2020}
{Wong} K.~C.,  et~al., 2020, \mn@doi [\mnras] {10.1093/mnras/stz3094}, \href
  {https://ui.adsabs.harvard.edu/abs/2020MNRAS.498.1420W} {498, 1420}

\bibitem[\protect\citeauthoryear{{Woods} et~al.,}{{Woods}
  et~al.}{2019}]{Woods_2019}
{Woods} T.~E.,  et~al., 2019, \mn@doi [\pasa] {10.1017/pasa.2019.14}, \href
  {https://ui.adsabs.harvard.edu/abs/2019PASA...36...27W} {36, e027}

\bibitem[\protect\citeauthoryear{{Wu} et~al.,}{{Wu} et~al.}{2015}]{Wu_2015}
{Wu} X.-B.,  et~al., 2015, \mn@doi [\nat] {10.1038/nature14241}, \href
  {http://adsabs.harvard.edu/abs/2015Natur.518..512W} {518, 512}

\bibitem[\protect\citeauthoryear{{Wyithe} \& {Loeb}}{{Wyithe} \&
  {Loeb}}{2002}]{Wyithe_Loeb_2002}
{Wyithe} J. S.~B.,  {Loeb} A.,  2002, \mn@doi [\nat] {10.1038/nature00794},
  \href {https://ui.adsabs.harvard.edu/#abs/2002Natur.417..923W} {417, 923}

\bibitem[\protect\citeauthoryear{{Yang} et~al.,}{{Yang}
  et~al.}{2016}]{Yang_2016}
{Yang} J.,  et~al., 2016, \mn@doi [\apj] {10.3847/0004-637X/829/1/33}, \href
  {https://ui.adsabs.harvard.edu/#abs/2016ApJ...829...33Y} {829, 33}

\bibitem[\protect\citeauthoryear{{Yue}, {Fan}, {Yang}  \& {Wang}}{{Yue}
  et~al.}{2021}]{Yue_2021_pair}
{Yue} M.,  {Fan} X.,  {Yang} J.,   {Wang} F.,  2021, \mn@doi [\apjl]
  {10.3847/2041-8213/ac31a9}, \href
  {https://ui.adsabs.harvard.edu/abs/2021ApJ...921L..27Y} {921, L27}

\bibitem[\protect\citeauthoryear{{Yue}, {Fan}, {Yang}  \& {Wang}}{{Yue}
  et~al.}{2022}]{Yue_2021}
{Yue} M.,  {Fan} X.,  {Yang} J.,   {Wang} F.,  2022, \mn@doi [\apj]
  {10.3847/1538-4357/ac409b}, \href
  {https://ui.adsabs.harvard.edu/abs/2022ApJ...925..169Y} {925, 169}

\makeatother
\end{thebibliography}





\bsp	
\label{lastpage}
\end{document}